\documentclass[onecolumn,journal,12pt,draftclsnofoot]{IEEEtran}
\usepackage[T1]{fontenc} 
\usepackage{cite}
\usepackage {graphicx}
\usepackage{amssymb}
\usepackage{amsmath}
\usepackage{algorithm}
\usepackage{algpseudocode}
\usepackage[caption=false, font=footnotesize]{subfig}
\usepackage{physics}
\usepackage[dvipsnames]{xcolor}
\usepackage{etoolbox}
\usepackage{hyperref}
\usepackage{multirow}
\usepackage{units}

\usepackage{stfloats}

\usepackage{flushend}

\begin{document}
\title{Deep Learning-Based Channel Estimation for Pinching Antenna-Enabled ISAC Systems}

\author{Alice~Faisal,~\IEEEmembership{Graduate~Member,~IEEE,}
    Mohamed~Elsayed,~\IEEEmembership{Member,~IEEE,}
Animesh~Yadav, \IEEEmembership{Senior~Member,~IEEE,} ~Ming~Zeng,~\IEEEmembership{Member,~IEEE,}
~Yuanwei~Liu,~\IEEEmembership{Fellow,~IEEE}
~and~Octavia~A.~Dobre,~\IEEEmembership{Fellow,~IEEE}%
       
}


\maketitle
\bstctlcite{IEEEexample:BSTcontrol}

\begin{abstract}
Pinching antennas (PAs) offer a flexible and energy-efficient architecture for integrated sensing and communication (ISAC) systems. However, the PA-induced propagation structure produces compressed, rank-deficient observations, making channel estimation challenging, especially with {limited pilot slots.} Consequently, classical estimators struggle due to the resulting underdetermined and ill-conditioned low-rank channel estimation problem. Motivated by this challenge, this work proposes a novel long short-term memory (LSTM)-based learning architecture in which the recurrent structure captures temporal dependencies in the received pilots and enables reliable  reconstruction of both communication and sensing channels despite the severe measurement compression. Building on the estimated sensing channel, a near-field multiple signal classification algorithm is further employed to recover target angle and range. The integration of learning-based channel reconstruction with model-based parameter extraction allows the proposed framework to preserve the geometric structure required for high-resolution sensing. Numerical evaluations demonstrate that the proposed LSTM estimator achieves robust and accurate channel reconstruction across a wide range of signal-to-noise ratios  and array configurations, including short-pilot and variable-PA settings. \textcolor{black}{Beyond outperforming classical estimation techniques, the architecture also offers substantial computational savings compared with state-of-the-art approaches, while achieving higher accuracy.} These results establish the proposed LSTM-driven framework as a practical and computationally efficient solution for future PA-enabled ISAC systems.
\end{abstract}


\IEEEpeerreviewmaketitle

 \vspace{-5mm}

\section{Introduction}



\IEEEPARstart{T}{he} {key performance indicators set for sixth-generation (6G) wireless networks will drive the development of use cases and applications that require unprecedented levels of massive connectivity, high spectral efficiency, and reliable coverage in increasingly dynamic and complex environments {\cite{6G_1}.} In such environments, maintaining dependable line-of-sight (LoS) connectivity remains a fundamental challenge.
To address this issue}, flexible-antenna technologies have been explored as a means to reconfigure wireless channels and enhance communication robustness {\cite{RIS_Principles, Fluid, Mov1}.} Although approaches such as reconfigurable intelligent surfaces, fluid antennas, and movable antennas  offer promising adaptability, their physical repositioning ranges are typically limited to the scale of a few wavelengths and thus remain insufficient for overcoming large-scale path loss or establishing {LoS links} in obstructed scenarios. These inherent constraints restrict their ability to deliver stable coverage and high-quality sensing performance in practical deployments. Motivated by these challenges, pinching antennas (PAs) have recently emerged as a novel dielectric-waveguide-based architecture that enables antenna activation at arbitrary positions along the waveguide with negligible cost and complexity \cite{Fukuda2022PinchingAntenna, PA_1}. This capability allows PAs to be relocated over distances spanning tens of meters, thereby significantly reducing propagation loss, improving LoS availability, and offering a highly scalable platform for flexible communication  in next-generation wireless systems \cite{PA_magazine,PA_2,PA_3}.

Additionally, integrated sensing and communication (ISAC) has gained substantial momentum as a key enabler for future wireless networks, driven by its ability to merge communication and radar sensing within a unified infrastructure \cite{ISAC_Opt_2023, Alice}. 
\textcolor{black}{This interest is also reflected in the 3rd Generation Partnership Project (3GPP) standardization process, where ISAC was introduced as a study topic in Release 19 and continues to be explored in Release 20 for emerging 6G use cases such as joint radar-communication, high-accuracy positioning, and environment-aware connectivity \cite{3gpp2016eutra}.}
{By allowing the same waveform, hardware, and spectrum to serve both data transmission and environmental sensing, ISAC enhances spectral efficiency, reduces hardware overhead, and supports emerging applications such as intelligent transportation, autonomous systems, extended reality, smart manufacturing, and digital-twin-enabled networks \cite{D1,D2,D3}.}
{However, the performance of ISAC systems is  influenced by the availability of reliable LoS links, which are particularly important for achieving accurate sensing while maintaining high-quality communication.}
In this context, integrating PAs with ISAC presents a highly promising direction: the wide-range deployability of PAs enables transceivers to be positioned closer to users or targets, thus mitigating large-scale path loss and enhancing the overall sensing-communication trade-off. This positions PA-enabled ISAC as a strong candidate for next-generation wireless architectures \cite{wijewardhana2025wirelessfedpinchingantennasystemswipass, PA_ISAC6}.

{Recent efforts have begun exploring the integration of PAs into ISAC systems, demonstrating the substantial performance gains achievable when combining flexible antenna positioning with joint sensing and communication (SAC) design \cite{PA_ISAC6, PA_ISAC2, PA_ISAC5, PA_ISAC1, Ding2025, PA_ISAC3, PA_ISAC7}}. \textcolor{black}{Existing studies have investigated PA-assisted ISAC from multiple angles, including beamforming optimization \cite{PA_ISAC2, PA_ISAC5}, power maximization \cite{PA_ISAC1}, and Cramér-Rao lower-bound characterization of positioning accuracy \cite{PA_ISAC6, Ding2025}.} These works collectively confirm that PAs can significantly enhance {the probability of LoS link}, improve echo strength, and yield superior communication-sensing trade-offs {by enabling  favorable propagation geometries for both data transmission and target sensing compared to conventional fixed antennas.} Despite this progress, the current literature on PA-enabled ISAC has primarily focused on system optimization under the assumption of perfect channel state information. While the works in \cite{PA_channel} and \cite{PA_channel2} have investigated channel estimation for PA-assisted communication systems, they focus solely on communication links. The former adopts a transformer-based uplink architecture, whereas the latter considers a \textcolor{black}{millimeter wave} setting with a classical estimation approach. To date, no existing work has addressed channel estimation in PA-enabled ISAC systems, where both SAC channels are acquired under the unique propagation conditions introduced by dielectric waveguides and flexible PA positioning. {This gap is particularly important as PA-enabled ISAC channel estimation is not a direct extension of conventional flexible-antenna channel estimation; rather, it requires recovering both SAC channels from pilot observations shaped jointly by the system geometry,  shared dielectric waveguide feed, and near-field SAC propagation environment.}

Channel estimation in PA-enabled ISAC is specifically challenging due to several inherent structural characteristics of PA architectures.  {Unlike conventional flexible-antenna systems, such as movable-antenna and fluid-antenna, where antenna reconfiguration mainly affects the free-space channel response, PA systems employ a single radio frequency (RF) feed per waveguide that supports multiple PAs. Consequently, the signals radiated from the activated PAs are not independently generated at each antenna location; instead, they are phase-shifted versions of the same injected waveform.} As a result, {the received pilot signal is} inherently low-dimensional, while the effective user and target channels are high-dimensional, leading to an intrinsically underdetermined and ill-conditioned recovery problem. Furthermore, the in-waveguide component is fully determined by the physical positions of the PAs on the waveguide, meaning that additional pilot transmissions do not generate new \textcolor{black}{independent observations} for disentangling the free-space and waveguide-induced effects. This limitation is further compounded by the flexible and dynamic placement of PAs, whose mobility causes the resulting wireless channels to exhibit strong near-field and non-stationary characteristics across both SAC links. Consequently, the estimation of SAC channels in PA-enabled ISAC systems requires fundamentally novel approaches that can cope with hybrid propagation, limited pilot diversity, and position-dependent channel behaviors.

{To this end, this work investigates channel estimation in PA-enabled ISAC systems, for the first time in the literature.} It considers a PA-assisted ISAC architecture comprising two dielectric waveguides: one dedicated to transmission and one to reception, thereby enabling simultaneous communication and target sensing. The formulation further incorporates environmental scatterers, reflecting the practical propagation conditions encountered in real deployments. To overcome these estimation challenges, a novel long short-term memory (LSTM)-based channel estimation framework specifically designed for PA-enabled ISAC systems and enriched with embedding-based feature representations is proposed. The presented model is capable of accurately recovering both SAC channels even when the number of pilot slots is smaller than the number of deployed PAs, a regime where conventional estimators fail due to the rank-deficient nature of the problem. Based on the reconstructed sensing channels, a two-dimensional (2D) near-field multiple signal classification (MUSIC) algorithm is further employed to extract target parameters, including range and angle, thus enabling high-resolution sensing for PA-enabled ISAC systems. Extensive evaluations demonstrate that the proposed approach achieves substantially higher accuracy and robustness than classical estimation techniques, highlighting its suitability for enabling reliable channel acquisition in practical PA-assisted ISAC systems. The contributions of this paper are summarized as follows:

\begin{itemize}
	\item A novel channel estimation framework is developed to estimate the SAC channels in PA-enabled ISAC systems. The formulation incorporates the  movement of PAs and the presence of environmental scatterers, ensuring that the resulting channel models reflect realistic propagation conditions in PA-assisted deployments.
	\item A dedicated feature representation is designed to enable effective learning-based channel estimation, and least-squares (LS) formulations are derived to serve as classical baseline estimators for both SAC channels.
	\item {A novel {LSTM-based} estimation model, specifically tailored for PA-enabled ISAC, is proposed.  The model aligns with the geometric structure of the PA-enabled system by combining temporal pilot processing, PA-location embeddings, and spatial recurrence along the  waveguide. This enables the estimator to capture  non-stationary behavior of PA-induced channels and achieve accurate reconstruction even when the number of pilot slots is smaller than the number of activated PAs.}
	\item A {2D-MUSIC-based} sensing \textcolor{black}{algorithm} is integrated to estimate target parameters using the reconstructed sensing channels. This facilitates high-resolution target parameter extraction compatible with the PA-enabled ISAC system.
	\item The computational complexity of the proposed framework is analyzed in terms of the number of real-valued additions and multiplications, providing insights into its scalability and suitability for real-time deployment.
	\item Extensive simulations across a broad range of signal-to-noise ratios (SNRs) and system configurations validate the effectiveness of the proposed approach. The results demonstrate that the proposed method achieves superior SAC channel estimation accuracy compared to conventional  schemes, while also reducing computational complexity compared to transformer-based approaches.
\end{itemize}


	The remainder of this paper is organized as follows: The system model and problem formulation {are introduced in  Section \ref{Sec:System_Model}}. The proposed {channel} estimation approach is {presented} in Section \ref{Sec:Proposed_Approach}, and the computational complexity is analyzed in Section \ref{Sec:Complexity_Analysis}. {Finally, simulation results are shown in Section \ref{Sec:Simulation_Results} and conclusions are drawn in Section \ref{Sec:Conclusion}.} 

    {Notations: Boldface uppercase and lowercase letters denote matrices and vectors, respectively. \(\mathbb{C}\) and  \(\mathcal{CN}\) stand for a complex-valued variable and a complex-valued normally distributed random variable, respectively.  \(\Re \{\cdot\}\) and \(\Im \{\cdot\}\) denote the real and imaginary components of a {complex variable}, respectively. \(\mathbb{E} \{\cdot\}\) denotes the expectation operation. The operators \( (\cdot)^{*}, (\cdot)^{\mathrm{H}}, (\cdot)^{\mathrm{T}},  \| \cdot \|\), and \(\| \cdot \|_F\) represent the complex conjugate, {conjugate transpose}, transpose, {Euclidean norm}, and Frobenius norm of their arguments, respectively. $\odot$ denotes element-wise multiplication. {The main notations used throughout the paper are summarized in Table~\ref{tab:notations}.}}

	\begin{table}[!t]
		\centering
		\caption{Summary of Key Notations}
		\label{tab:notations}
		\scriptsize
		\renewcommand{\arraystretch}{0.9}
		{
		\begin{tabular}{p{0.13\columnwidth}p{0.25\columnwidth}|
				p{0.13\columnwidth}p{0.25\columnwidth}}
			\hline
			\textbf{Notation} & \textbf{Description} &
			\textbf{Notation} & \textbf{Description} \\
			\hline
			$K$, $N$ & UEs, transmit PAs &
			$T$, $S$ & Pilot slots, scatterers \\
			$D_x$, $D_y$, $d$ & Region size, height &
			$\mathcal X$ & PA candidate set \\
			$\boldsymbol{\psi}_k$ & UE-$k$ position &
			$\boldsymbol{\psi}_{\rm tar}$ & Target position \\
			$\boldsymbol{\psi}^{\rm tx}_n$ & Transmit-PA position &
			$\boldsymbol{\psi}^{\rm rx}$ & Receive-PA position \\
			$\mathbf g_{\rm tx}$ & Tx waveguide  &
			$g_{\rm rx}$ & Rx waveguide  \\
			$\mathbf h_{u,k}$ & UE-$k$ channel &
			$\boldsymbol{\Theta}_{\rm LoS}$ & LoS sensing channel \\
			$\boldsymbol{\Theta}_{\rm NLoS}$ & Sensing clutter &
			$\beta$, $\beta_s$ & Reflection coefficients \\
			$y_{k,t}$ & UE pilot signal &
			$z_t$ & BS sensing echo \\
			$\mathbf R_{\rm UE_k}$ & UE NN input &
			$\mathbf R_{\rm BS}$ & Sensing NN input \\
			$\mathbf O_{\rm UE_k}$ & UE NN label &
			$\mathbf O_{\rm BS}$ & Sensing NN label \\
			$\tilde{\mathbf x}^{\rm tx}$ & Tx PA positions &
			$\tilde{\mathbf x}^{\rm rx}$ & Rx PA position \\
			$\phi(\cdot)$, $L_f$ & Fourier embedding &
			$\mathcal L_{\rm MAE}$ & Training loss \\
			$\mathbf u_n(\tau)$ & PA-wise input &
			$\mathbf f_n$ & Temporal feature \\
			$\mathbf t_n$ & Fused feature &
			$\mathbf s_n$ & Spatial feature \\
			$\hat{\mathbf O}_{\rm UE_k}$ & UE NN output &
			$\hat{\mathbf O}_{\rm BS}$ & Sensing NN output \\
			$r$, $\varphi$ & Range, azimuth &
			$\mathbf a(r,\varphi)$ & Steering vector \\
			$\hat{\mathbf R}_{\theta}$ & MUSIC covariance &
			$\mathbf E_n$ & Noise subspace \\
			$\mathcal G$ & MUSIC grid &
			$P_{\rm MUSIC}$ & MUSIC spectrum \\
			\hline
		\end{tabular}}
	\end{table}


\vspace{5mm}
	
\section{System Model}~\label{Sec:System_Model}



\begin{figure}[!t]
		\centering
		\includegraphics[width=0.6\textwidth]{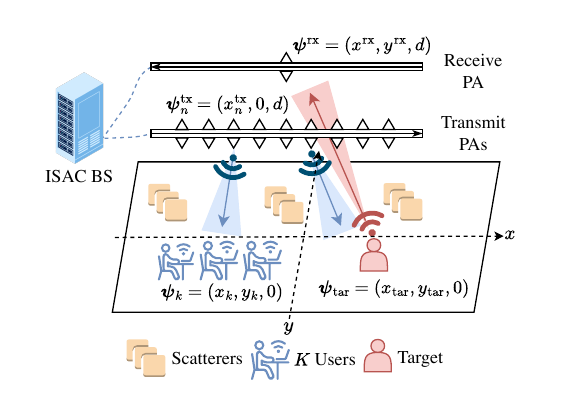}
		\caption{PA-enabled ISAC system.} 
		\label{Fig:sys}
\end{figure}

Consider a PA-enabled ISAC system, where a dual-function base station (BS) simultaneously serves multiple {single-antenna user equipments (UEs) in the downlink channel} while sensing a target, as shown in Fig. \ref{Fig:sys}. 
{The BS is connected to two dielectric waveguides, implementing an ISAC architecture with separate transmission and reception paths.}
The first waveguide is dedicated to transmission and is equipped with $N$ PAs, which radiate the downlink SAC signals. The second waveguide is dedicated to reception and is equipped with a single PA, which captures the \textcolor{black}{echo signal} reflected from the target and forwards it to the BS for {target parameter estimation (i.e., range and azimuth angle).} To better reflect practical deployment scenarios, multiple environmental scatterers are assumed to be present in the coverage region. These scatterers contribute additional multipath components to {both the downlink communication channels and received sensing echoes at the BS.} As a result, the received signals consist of  LoS components and scattered non-LoS (NLoS) components, leading to a more realistic propagation model for PA-enabled ISAC systems.

A three-dimensional (3D) Cartesian coordinate system is considered to describe the PA-enabled ISAC system. The coverage region lies on the $x$-$y$ plane and is modeled as a rectangular area with dimensions $D_x$ and $D_y$. {A set of $K$ single-antenna downlink UEs is uniformly distributed within this region, and the position of the $k$th UE is denoted by {$\boldsymbol{\psi}_k = (x_k, y_k, 0),  k = 1, \dots, K$}.}
In addition, the sensing target is located within the same region, whose position is represented by $	\boldsymbol{\psi}_{\mathrm{tar}} = (x_{\mathrm{tar}}, y_{\mathrm{tar}}, 0)$. {As in \cite{PA_ISAC1, PA_ISAC2, PA_ISAC3}, the target is assumed to be static or moving at a sufficiently low speed such that the Doppler effect is neglected over the considered pilot acquisition interval.} The two dielectric waveguides are assumed to be parallel to the $x$-axis {and mounted} at a fixed height $d$ {(e.g., ceiling)} above the $x$-$y$ plane. To this end, {the positions of the transmit PAs  are denoted by} $\boldsymbol{\psi}_n^{\mathrm{tx}} = (x_n^{\mathrm{tx}}, 0, d), \quad n = 1, \dots, N$, whereas the receive PA is positioned at $\boldsymbol{\psi}^{\mathrm{rx}} = (x^{\mathrm{rx}}, y^{\mathrm{rx}}, d).$ Here, $x_n^{\mathrm{tx}}, x^{\mathrm{rx}} \in [0, D_x]$, where $D_x$ represents the horizontal span of the waveguide.

In this work, a discrete PA activation strategy is adopted to reflect practical hardware constraints. Specifically, PAs can only be deployed at predefined discrete locations along the transmitting waveguide \cite{PA_channel}. Let the set of feasible activation positions be defined as
\begin{equation}
	\mathcal{X} = \left\{ \frac{D_x}{Q-1}(q-1) \,\bigg|\, q = 1, 2, \dots, Q \right\},
\end{equation}
where $Q$ denotes the total number of candidate discrete locations. From this feasible set, $N$ PAs are randomly selected and activated for transmission. {These $N$ activated PAs jointly serve the downlink UEs and the sensing task; hence, each UE receives the aggregate signal radiated by all active transmit PAs.} Since all transmit PAs are mounted on the same dielectric waveguide, the signals radiated from different PAs originate from a common feed point and propagate along the waveguide before being coupled into free space. Consequently, the signal emitted from each PA is a phase-shifted version of the injected signal, where the phase shift depends on the guided propagation distance between the feed point and the corresponding PA location. This structure inherently couples the transmitted signals across PAs and introduces deterministic phase progression governed by the waveguide properties. Let $\boldsymbol{\psi}^{\mathrm{tx}}_{0}$ and $\boldsymbol{\psi}^{\mathrm{rx}}_{0}$ denote the excitation (feed) points of the transmitting and receiving waveguides, respectively. The in-waveguide channel from the transmit feed point to the $n$th transmitting PA located at $\boldsymbol{\psi}^{\mathrm{tx}}_{n}$ can be expressed as
\begin{equation}
	{g}_{\mathrm{tx},n}
	= \sqrt{\frac{P_{\mathrm{tx}}}{N}}
	\exp\!\left(
	-j k_{w}
	\left\|
	\boldsymbol{\psi}^{\mathrm{tx}}_{0}
	-
	\boldsymbol{\psi}^{\mathrm{tx}}_{n}
	\right\|
	\right),
	\quad n = 1, \ldots, N,
\end{equation}
where $P_{\mathrm{tx}}$ denotes the total transmit power injected into the transmitting waveguide. The factor $\sqrt{P_{\mathrm{tx}}/N}$ ensures equal power allocation across active transmit PAs. {The guided wavenumber is defined as $k_w = 2\pi/\lambda_w$, equivalently $k_w = 2\pi\eta_{\mathrm{eff}}/\lambda$}, {where $\lambda$ is the free-space wavelength corresponding to the carrier frequency $f_c$,} $\lambda_{w} = \lambda/\eta_{\mathrm{eff}}$ is the guided wavelength inside the dielectric waveguide, and $\eta_{\mathrm{eff}}$ denotes its effective refractive index \cite{PA_Opt}. The corresponding in-waveguide channel vector is defined as $\mathbf{g}_{\mathrm{tx}}
	=
	\left[
	g_{\mathrm{tx},1}, \ldots, g_{\mathrm{tx},N}
	\right]^{\mathsf T}
	\in \mathbb{C}^{N \times 1}.$ Similarly, the in-waveguide channel from the receiving PA located at $\boldsymbol{\psi}^{\mathrm{rx}}$ to the receive feed point $\boldsymbol{\psi}^{\mathrm{rx}}_{0}$ is modeled as $
		g_{\mathrm{rx}}
		=
		\exp\!\left(
		-j k_{w}
		\left\|
		\boldsymbol{\psi}^{\mathrm{rx}}
		-
		\boldsymbol{\psi}^{\mathrm{rx}}_{0}
		\right\|
		\right).$

  For the propagation environment, a spherical wavefront  model is adopted to accurately characterize the near-field channel between the transmit PAs and the downlink UEs. Based on this, the LoS channel between the transmitting PAs and the $k$th UE, $\mathbf{h}_{u,k}^{\mathrm{LoS}} \in \mathbb{C}^{N \times 1}$, is expressed as

%

\begin{equation}
	\mathbf{h}_{u,k}^{\mathrm{LoS}}
	=
	\left[
	\frac{\sqrt{\eta}\,
		e^{-j \frac{2\pi}{\lambda}
			\left\|
			\boldsymbol{\psi}_k -
			\boldsymbol{\psi}^{\mathrm{tx}}_{1}
			\right\|}}
	{\left\|
		\boldsymbol{\psi}_k -
		\boldsymbol{\psi}^{\mathrm{tx}}_{1}
		\right\|},
	\;
	\ldots,
	\;
	\frac{\sqrt{\eta}\,
		e^{-j \frac{2\pi}{\lambda}
			\left\|
			\boldsymbol{\psi}_k -
			\boldsymbol{\psi}^{\mathrm{tx}}_{N}
			\right\|}}
	{\left\|
		\boldsymbol{\psi}_k -
		\boldsymbol{\psi}^{\mathrm{tx}}_{N}
		\right\|}
	\right]^{\mathsf T},
\end{equation}

\noindent {where  $\eta = \lambda^2/(16\pi^2)$ represents the reference path-loss factor.} 


{In realistic deployment scenarios, the propagation environment typically includes multiple surrounding objects, such as building surfaces, walls, and metallic structures, that introduce additional scattered components.} \textcolor{black}{Let $S$ denote} the number of scatterers within the coverage region. The resulting NLoS component of the UE channel can be modeled as
\begin{equation}
	\mathbf{h}_{u,k}^{\mathrm{NLoS}}
	=
	\sqrt{\frac{1}{S}}
	\sum_{s=1}^{S}
	\alpha_{k,s}
	\,\mathbf{v}_{k,s},
\end{equation}
where $\alpha_{k,s} \sim \mathcal{CN}(0,\sigma_s^2)$ represents the complex gain associated with the $s$th scattering path. The array response corresponding to the {$s$th} scatterer is characterized using a spherical-wave model and expressed as \cite{PA_channel}
\begin{equation}
	\label{eq:user_scatter}
	\mathbf{v}_{k,s}
	=
	\left[
	\frac{\sqrt{\eta}\,
		e^{-j \frac{2\pi}{\lambda}
			(d_{k,s} + d_{s,1})}}
	{d_{k,s}\, d_{s,1}},
	\;
	\ldots,
	\;
	\frac{\sqrt{\eta}\,
		e^{-j \frac{2\pi}{\lambda}
			(d_{k,s} + d_{s,N})}}
	{d_{k,s}\, d_{s,N}}
	\right]^{\mathsf T},
\end{equation}
where $d_{k,s}$ denotes the distance between UE $k$ and scatterer $s$, and $d_{s,n}$ represents the distance from scatterer $s$ to the {$n$th} transmit PA. Accordingly, the overall communication channel of UE $k$ is given by
$
	\mathbf{h}_{u,k}
	=
	\boldsymbol{\zeta}_k \odot \mathbf{h}_{u,k}^{\mathrm{LoS}}
	+
	\mathbf{h}_{u,k}^{\mathrm{NLoS}},
$
where $\boldsymbol{\zeta}_k = [\zeta_{k,1},\ldots,\zeta_{k,N}]^{\mathsf T}$, is a Bernoulli vector, and $\zeta_{k,n}$ indicates the existence of a LoS path between UE $k$ and PA $n$.

 Similarly, the LoS channel between the transmitting PAs and the target, \textcolor{black}{$\mathbf{h}_{\mathrm{tar},\mathrm{tx}} \in \mathbb{C}^{N\times 1}$, is given by}
\begin{equation}
	\mathbf{h}_{\mathrm{tar},\mathrm{tx}}
	=
	\left[
	\frac{\sqrt{\eta}\,
		e^{-j \frac{2\pi}{\lambda}\left\|
			\boldsymbol{\psi}_{\mathrm{tar}}-\boldsymbol{\psi}^{\mathrm{tx}}_{1}
			\right\|}}
	{\left\|
			\boldsymbol{\psi}_{\mathrm{tar}}-\boldsymbol{\psi}^{\mathrm{tx}}_{1}
		\right\|},
	\;
	\ldots,
	\;
	\frac{\sqrt{\eta}\,
		e^{-j \frac{2\pi}{\lambda}\left\|
			\boldsymbol{\psi}_{\mathrm{tar}}-\boldsymbol{\psi}^{\mathrm{tx}}_{N}
			\right\|}}
	{\left\|
		\boldsymbol{\psi}_{\mathrm{tar}}-\boldsymbol{\psi}^{\mathrm{tx}}_{N}
		\right\|}
	\right]^{\mathsf T}.
\end{equation}
Moreover, the LoS channel from the target to the receive PA is modeled as
\begin{equation}
	h_{\mathrm{tar},\mathrm{rx}}
	=
	\frac{\sqrt{\eta}\,
		e^{-j \frac{2\pi}{\lambda}\left\|
				\boldsymbol{\psi}_{\mathrm{tar}}-\boldsymbol{\psi}^{\mathrm{rx}}
			\right\|}}
	{\left\|
			\boldsymbol{\psi}_{\mathrm{tar}}-\boldsymbol{\psi}^{\mathrm{rx}}
		\right\|}
	\in \mathbb{C}.
\end{equation}

\noindent To model sensing clutter, the received sensing link is further assumed to include NLoS components generated by $S$ environmental scatterers. In contrast to the downlink UE channel, each sensing NLoS contribution experiences a cascaded propagation from the transmit PAs to the scatterer and  from the scatterer to the receive PA. Hence, the NLoS sensing channel can be represented by an $N\times 1$ vector whose {$n$th entry} follows the same spherical-wave structure as in \eqref{eq:user_scatter}.


{For channel estimation, the BS transmits pilot symbols to the UEs over $T$ time slots. Let $t=1,\ldots,T$ denote the pilot slot index. The received signal at UE $k$ during the $t$th pilot slot is given by}
\begin{equation}
	\label{eq:user}
	y_{k,t}
	=
	\mathbf{h}_{u,k}^{\mathsf H}\,\mathbf{g}_{\mathrm{tx},t}\,p_t
	+
	n_{k,t},
\end{equation}
{where $p_t=\exp\!\left(-j\pi(t-1)^2/T\right)$ denotes the transmitted pilot symbol, and ${n}_{k,t} \sim \mathcal{CN}(0, \sigma_{k,t}^2)$ represents the additive white complex Gaussian noise (AWGN) with zero mean and variance of $\sigma_{k,t}^2$.} On the other hand, the echo {(reflected)} signal received at the BS during the {$t$th} pilot slot is given by
\begin{equation}
	\label{eq:target}
	\begin{aligned}
		z_t
		&=
		\underbrace{
			 g_{\mathrm{rx},t}^{*}
			\Big(
			\beta \, h_{\mathrm{tar},\mathrm{rx}} \, \mathbf{h}_{\mathrm{tar},\mathrm{tx}}^{\mathsf H}
			\Big)
			\mathbf g_{\mathrm{tx},t}\, p_t
		}_{\text{LoS target return}}
		\\
		&\quad+
		\underbrace{
			 g_{\mathrm{rx},t}^{*}
			\left(
			\frac{1}{\sqrt{S}}
			\sum_{s=1}^{S}
			\beta_s \, h_{s,\mathrm{rx}} \, \mathbf h_{s,\mathrm{tx}}^{\mathsf H}
			\right)
			\mathbf g_{\mathrm{tx},t}\, p_t
		}_{\text{NLoS clutter}}
		+
		n_t',
	\end{aligned}
\end{equation}
where $\beta$ and  $\beta_s$ denote the complex reflection coefficient of the target and the {$s$th} scatterer, respectively. {$\mathbf h_{s,\mathrm{tx}}$ and $h_{s,\mathrm{rx}}$ are the free-space propagation channels from the transmit PAs to scatterer $s$, and from scatterer $s$ to the receive PA, respectively, modeled using the same spherical-wave propagation structure as in \eqref{eq:user_scatter}.}
$n_t' \sim \mathcal{CN}(0,\sigma_t^2)$ {represents
AWGN at the receiver.} 
{As in \cite{interference_PA2}, it is assumed that the PAs are sufficiently spaced, such that mutual coupling is negligible. Specifically, the minimum inter-PA spacing is approximately $9.4\lambda$, exceeding the $\lambda/2$ separation considered in \cite{interference_PA2}.}




For notational convenience, the effective sensing channel corresponding to the LoS target is defined as
\begin{equation}
\label{eq:theta}
	\boldsymbol{\Theta}_{\mathrm{LoS}}
	\triangleq
	\beta \,
	h_{\mathrm{tar},\mathrm{rx}}
	\, \mathbf h_{\mathrm{tar},\mathrm{tx}}^{\mathsf H}
	\in \mathbb{C}^{1 \times N}.
\end{equation}

\noindent Similarly, the aggregate NLoS sensing component induced by $S$ environmental scatterers is modeled as
\begin{equation}
	\boldsymbol{\Theta}_{\mathrm{NLoS}}
	=
	\frac{1}{\sqrt{S}}
	\sum_{s=1}^{S}
	\beta_s \,
	h_{s,\mathrm{rx}}
	\, \mathbf h_{s,\mathrm{tx}}^{\mathsf H} \in \mathbb{C}^{1 \times N}.
\end{equation}

\noindent Accordingly, \eqref{eq:target} can be compactly expressed as
\begin{equation}
	\label{eq:target_f}
	z_t
	=
	 g_{\mathrm{rx},t}^{*}
	\left(
	\boldsymbol{\Theta}_{\mathrm{LoS}}
	+
	\boldsymbol{\Theta}_{\mathrm{NLoS}}
	\right)
	\mathbf g_{\mathrm{tx},t}
	\, p_t
	+
	n_t'.
\end{equation}
\textit{Channel Estimation Problem:}
\textcolor{black}{The objective} of this work is to accurately estimate the effective communication channel $\mathbf{h}_{u,k}$ and the LoS sensing channel associated with the target. However, this estimation problem is inherently challenging in PA-enabled ISAC systems. Due to the in-waveguide signal coupling, the received observations at both the UEs and the BS collapse into scalar measurements per pilot slot, while the underlying SAC channels remain high-dimensional vectors. Consequently, high-dimensional channel parameters must be recovered from low-dimensional observations. 
{Notably, even if $T \geq N$, activating all PAs simultaneously does not provide $N$ independent observations, leading to a compressed measurement model.}
The sensing channel estimation is further complicated by the bistatic round-trip propagation structure, where the target return depends on the cascaded effect of the transmitting and receiving waveguides. {Moreover, the sensing parameter of interest corresponds to the LoS target channel rather than the aggregate sensing response, whereas the NLoS components constitute clutter in the received observations \cite{ISAC_clutter}.} These structural characteristics render conventional estimation techniques inadequate. Therefore, structurally aware estimation frameworks are required, which motivates the development of the novel channel estimation approach presented in the next section.

\section{Proposed Estimation Approach}~\label{Sec:Proposed_Approach}	

{The channel estimation problem in PA-enabled ISAC systems cannot be directly addressed using conventional LS techniques due to the inherent signal coupling among PAs introduced by the shared waveguide feed.}
In particular, without structural changes, the received signals across pilot slots do not provide sufficiently independent observations to uniquely recover the high-dimensional channel vectors. To enable a feasible baseline, {a PA switching measurement scheme is adopted}, where controlled {PA activation} patterns are deployed across pilot slots to generate linearly independent observations. While this approach establishes a classical benchmark, it remains fundamentally limited \textcolor{black}{in  exploiting} the structured propagation characteristics and nonlinear dependencies present in PA-assisted ISAC systems. To overcome this, a dedicated learning-based estimation framework is proposed. The details of the switching-based LS baseline and the proposed deep learning estimator are presented in the following subsections.
	\subsection{Channel Estimation Framework}
	
		\subsubsection{{PA} Switching-Based LS Baseline Estimation} \label{sec:LS}
	
		Considering the received signal in \eqref{eq:user}, direct application of LS is not feasible under simultaneous activation of all PAs. In this case, each pilot slot yields only a single scalar observation, which is insufficient to uniquely recover the $N$-dimensional channel vector, $\mathbf h_{u,k}$. To generate linearly independent measurements, a sequential {PA} switching scheme is \textcolor{black}{adopted \cite{PA_channel}}. Let $\mathbf w_t \in \mathbb{C}^{N\times 1}$ \textcolor{black}{denote} a binary activation (i.e., masking) vector applied during the {$t$th} pilot slot. In the adopted switching strategy, only one PA is activated per slot, such that
		\begin{equation}
		\mathbf w_t \in \{0,1\}^{N}, \quad
		[\mathbf w_t]_n =
		\begin{cases}
			1, & n=t, \\
			0, & \text{otherwise}.
		\end{cases}
	\end{equation}

		 Consequently, $T = N$ pilot slots are required to sequentially probe all PAs. Under this switching mechanism, the effective transmit waveguide vector becomes
		$
			\mathbf g_{\mathrm{tx},t}
			=
			\mathbf g_{\mathrm{tx}} \odot \mathbf w_t,
		$
		where $\odot$ denotes the Hadamard product. Accordingly, the received signal at UE $k$ in slot $t$ can be written as
		\begin{equation}
			y_{k,t}
			=
			\mathbf h_{u,k}^{\mathsf H}
			\big( \mathbf g_{\mathrm{tx}} \odot \mathbf w_t \big)
			p_t
			+
			n_{k,t}.
		\end{equation}
		
		\noindent Stacking the received signals { at UE $k$} over $T$ pilot slots yields
		\begin{equation}
			\label{eq:user_s}
			\mathbf y_k
			=
			\mathbf A_k \mathbf h_{u,k}
			+
			\mathbf n_k,
		\end{equation}
		where $	\mathbf y_k = \left[y_{k,1}, \cdots, y_{k,T}\right]^{\mathsf T} \in \mathbb{C}^{T\times 1}$, $\mathbf n_k = \left[n_{k,1}, \cdots, n_{k,T}\right]^{\mathsf T} \in \mathbb{C}^{T\times 1}$, and the measurement matrix $\mathbf A_k \in \mathbb{C}^{T\times N}$ is defined as
		\begin{equation}
			\mathbf A_k
			=
			\begin{bmatrix}
				p_1 (\mathbf g_{\mathrm{tx}} \odot \mathbf w_1)^{\mathsf H} \\
				\vdots \\
				p_T (\mathbf g_{\mathrm{tx}} \odot \mathbf w_T)^{\mathsf H}
			\end{bmatrix}.
		\end{equation}
		
		\noindent  Thus, the LS estimate of the {$k$th UE} channel is given by
		\begin{equation}
			\hat{\mathbf h}_{u,k}^{\mathrm{LS}}
			=
			\left( \mathbf A_k^{\mathsf H} \mathbf A_k \right)^{-1}
			\mathbf A_k^{\mathsf H} \mathbf y_k.
		\end{equation}

		Similarly, stacking the received sensing signals { at the BS} in \eqref{eq:target_f} over $T$ pilot slots yields

		\begin{equation}
			\mathbf z
			=
			\mathbf A_s
			\left(
			\boldsymbol{\Theta}_{\mathrm{LoS}}^{\mathsf H}
			+
			\boldsymbol{\Theta}_{\mathrm{NLoS}}^{\mathsf H}
			\right)
			+
			\mathbf n',
		\end{equation}
		where $\mathbf z = [z_1, \ldots, z_T]^{\mathsf T} \in \mathbb{C}^{T \times 1}$  and $\mathbf n' = [n'_1, \ldots, n'_T]^{\mathsf T} \in \mathbb{C}^{T \times 1}$. Under the same sequential switching scheme adopted for communication channel estimation, the sensing measurement matrix $\mathbf A_s \in \mathbb{C}^{T \times N}$ is given by
		\begin{equation}
			\mathbf A_s
			=
			\begin{bmatrix}
				p_1 g_{\mathrm{rx},1}^{*}
				(\mathbf g_{\mathrm{tx}} \odot \mathbf w_1)^{\mathsf H}
				\\
				\vdots
				\\
				p_T g_{\mathrm{rx},T}^{*}
				(\mathbf g_{\mathrm{tx}} \odot \mathbf w_T)^{\mathsf H}
			\end{bmatrix}.
		\end{equation}
		
		\noindent \textcolor{black}{Based on this,} the LS estimate of the LoS sensing channel can be expressed as
		\begin{equation}
			\label{eq:LS_target}
			\hat{\boldsymbol{\theta}}^{\mathrm{LS}}
			=
			\left(
			\mathbf A_s^{\mathsf H} \mathbf A_s
			\right)^{-1}
			\mathbf A_s^{\mathsf H}
			\mathbf z. 
		\end{equation}
		{The LS baseline assumes knowledge of the pilot symbols, PA switching patterns, and deterministic in-waveguide responses determined by the PA locations and waveguide parameters.}
		It is important to emphasize that the sensing observations inherently contain the additional structured interference induced by $\boldsymbol{\Theta}_{\mathrm{NLoS}}$. As a result, even at high SNR, the {LS estimator remains inaccurate} due to the presence of clutter. It is also worth noting that both the SAC LS estimators require $T=N$ pilot slots to obtain a full-rank measurement matrix. Moreover, their performance deteriorates significantly in the presence of noise, as they fail to exploit the structured coupling induced by the waveguide architecture and the underlying propagation geometry. These limitations motivate the development of a structurally aware learning-based estimation framework, which is {presented next.}

		
		\subsubsection{{SAC Feature Representation Design}}

		\paragraph{{Communication Link}} For the learning-based estimator, full activation of the transmitting \textcolor{black}{PAs is} considered during the pilot phase. In this case, the received pilot observations at UE $k$ are collected across $T$ pilot symbols as defined in \eqref{eq:user_s}. To enhance the spatial structure of the observations, the received pilot sequence is projected {onto the waveguide  propagation structure.} Specifically,  define
		\begin{equation}
			\mathbf{Y}_{k}
			=
			\mathbf{g}_{\mathrm{tx}} \,
			\mathbf{y}_{k}^{\mathsf T}
			\in
			\mathbb{C}^{N \times T},
		\end{equation}
		where each column corresponds to a pilot observation weighted by the deterministic waveguide propagation structure. Accordingly, the communication input feature matrix is constructed by concatenating the normalized PA locations, the real and imaginary parts of $\mathbf{Y}_{k}$, and the pilot symbols. {Let $\mathbf{x}^{\mathrm{tx}} = [x_1^{\mathrm{tx}},\ldots,x_N^{\mathrm{tx}}]^{\mathsf T}$, the normalized antenna position vector is defined as $\tilde{\mathbf{x}}^{\mathrm{tx}} = {\mathbf{x}^{\mathrm{tx}}}/{D_x}.$ The final communication input {representation} is given by}
		\begin{equation}
			\label{eq:user_input}
			\begin{split}			
			\mathbf{R}^{\text{UE}_k}
			&=
			[
			\tilde{\mathbf{x}}^{\mathrm{tx}} ,\;
			\Re\{\mathbf{Y}_{k}\},\;
			\Im\{\mathbf{Y}_{k}\},\; \\
			& \Re\{\mathbf{1}_N \mathbf{p}^{\mathsf T}\},\;
			\Im\{\mathbf{1}_N \mathbf{p}^{\mathsf T}\}
			] \in
			\mathbb{R}^{N \times (1+4T)},
		\end{split}
		\end{equation}
		\textcolor{black}{where $\mathbf{p} = [p_1,\ldots,p_T]^{\mathsf{T}} \in \mathbb{C}^{T \times 1}$ denotes the pilot vector, and $\mathbf{1}_N \in \mathbb{R}^{N \times 1}$ is an all-ones column vector.}
		 The corresponding output (i.e., ground-truth label) is the effective communication channel vector
		\begin{equation}
			\mathbf{O}^{\text{UE}_k}
			=
			\left[
			\Re\{\mathbf{h}_{u,k}\},\;
			\Im\{\mathbf{h}_{u,k}\}
			\right]
			\in
			\mathbb{R}^{N \times 2}.
		\end{equation}

		\paragraph{{Sensing Link}}For sensing, the BS observes the scalar echo signal in \eqref{eq:target} over $T$ pilot slots. Compared to the communication link, the sensing return exhibits a bistatic round-trip structure governed by the cascaded transmit/receive waveguide responses as well as the target reflection coefficient, while the observation is additionally affected by the clutter term $\boldsymbol{\Theta}_{\mathrm{NLoS}}$. Moreover, the parameter of interest is not the aggregate effective sensing channel,  $\boldsymbol{\Theta}_{\mathrm{LoS}}+\boldsymbol{\Theta}_{\mathrm{NLoS}}$, but rather the {LoS target channel} $\boldsymbol{\Theta}_{\mathrm{LoS}}$ {defined in \eqref{eq:theta}}. This makes the estimation task intrinsically more challenging, since $\boldsymbol{\Theta}_{\mathrm{NLoS}}$ cannot be treated as simple AWGN and its effect does not vanish {even in low-noise regimes.}
        Therefore, richer feature construction is adopted for learning-based sensing estimation.
		
		To this end, the sensing feature construction incorporates multiple sources of structural information. First, the stacked sensing observations $\mathbf z=[z_1,\ldots,z_T]^{\mathsf T}$ are included and lifted to align the scalar measurements with the $N$-dimensional channel structure. Second, since the receive-PA location directly affects the target-to-receiver propagation path, the normalized receive-PA position is  included \textcolor{black}{as an  input feature} in addition to $\tilde{\mathbf{x}}^{\mathrm{tx}}$. Third, to provide a coarse initialization, the switching-based LS estimate obtained in \eqref{eq:LS_target} is appended as an auxiliary feature, enabling the learning model to refine a physics-based baseline rather than estimating the LoS component from scratch. {Let
		$\tilde{\mathbf{x}}^{\mathrm{rx}}=({x^{\mathrm{rx}}}/{D_x})\mathbf 1_N\in\mathbb{R}^{N\times 1}$ denote the normalized receive-PA vector. Based on this, the final sensing input matrix is constructed as}

		\begin{equation}
			\label{eq:target_input}
			\begin{split}
			\mathbf{R}^{\text{BS}}
			&=
			\big[\tilde{\mathbf{x}}^{\mathrm{tx}},\;
			\tilde{\mathbf{x}}^{\mathrm{rx}},\;
			\Re\{\mathbf 1_N\mathbf z^{\mathsf T}\},\;
			\Im\{\mathbf 1_N\mathbf z^{\mathsf T}\},\; \\ &
			\Re\{\mathbf 1_N\mathbf p^{\mathsf T}\},\;
			\Im\{\mathbf 1_N\mathbf p^{\mathsf T}\},\; \\ &
			\Re\{\hat{\boldsymbol{\theta}}^{\mathrm{LS}}\},\;
			\Im\{\hat{\boldsymbol{\theta}}^{\mathrm{LS}}\}\big]
			\in
			\mathbb{R}^{N\times (4+4T)}.
		\end{split}
		\end{equation}
		
		The corresponding output is the LoS sensing channel vector associated with the target, defined as
		\begin{equation}
        \label{eq:outputBS}
			\mathbf{O}^{\text{BS}}
			=
			\left[
			\Re\{\boldsymbol{\Theta}_{\mathrm{LoS}}^{\mathsf T}\},\;
			\Im\{\boldsymbol{\Theta}_{\mathrm{LoS}}^{\mathsf T}\}
			\right]
			\in
			\mathbb{R}^{N\times 2}.
		\end{equation}

		\noindent Given the input-output design for both SAC links, the following sub-section details the proposed network for solving the estimation problem.
		
		
		\subsubsection{Deep Learning-Based LSTM Estimator}
		\label{sec:DLEst}
		
		The objective of the proposed neural network (NN) is to learn a nonlinear mapping from the {structured pilot observations} to the underlying high-dimensional channel vectors. In the considered PA-enabled ISAC system, the received signals over $T$ pilot slots exhibit two key structural properties: (i) a temporal structure across pilot symbols governed by the transmitted sequence, and (ii) a spatial structure across the $N$ transmit PAs induced by the deterministic waveguide propagation and geometry-dependent channel variations. {Unlike generic flexible-antenna systems, this spatial structure arises not only from free-space position-dependent propagation, but also  from the common-feed dielectric waveguide, which imposes deterministic in-waveguide phase progression across the activated PAs. Therefore, the estimation problem is inherently spatio-temporal. Conventional fully connected NNs treat the input as an unstructured vector, thereby ignoring the sequential nature of pilot observations and the ordered spatial correlation across PAs.} 
		{In contrast, recurrent architectures are specifically designed to capture dependencies along ordered dimensions. Among these, LSTM offers a more flexible memory mechanism than simpler recurrent architectures through its separate gated cell state, enabling selective retention of informative pilot-domain features under compressed and clutter-contaminated observations.}

\begin{figure*}[!t]
        \makebox[\textwidth][l]{%
        \includegraphics[width=1\textwidth]{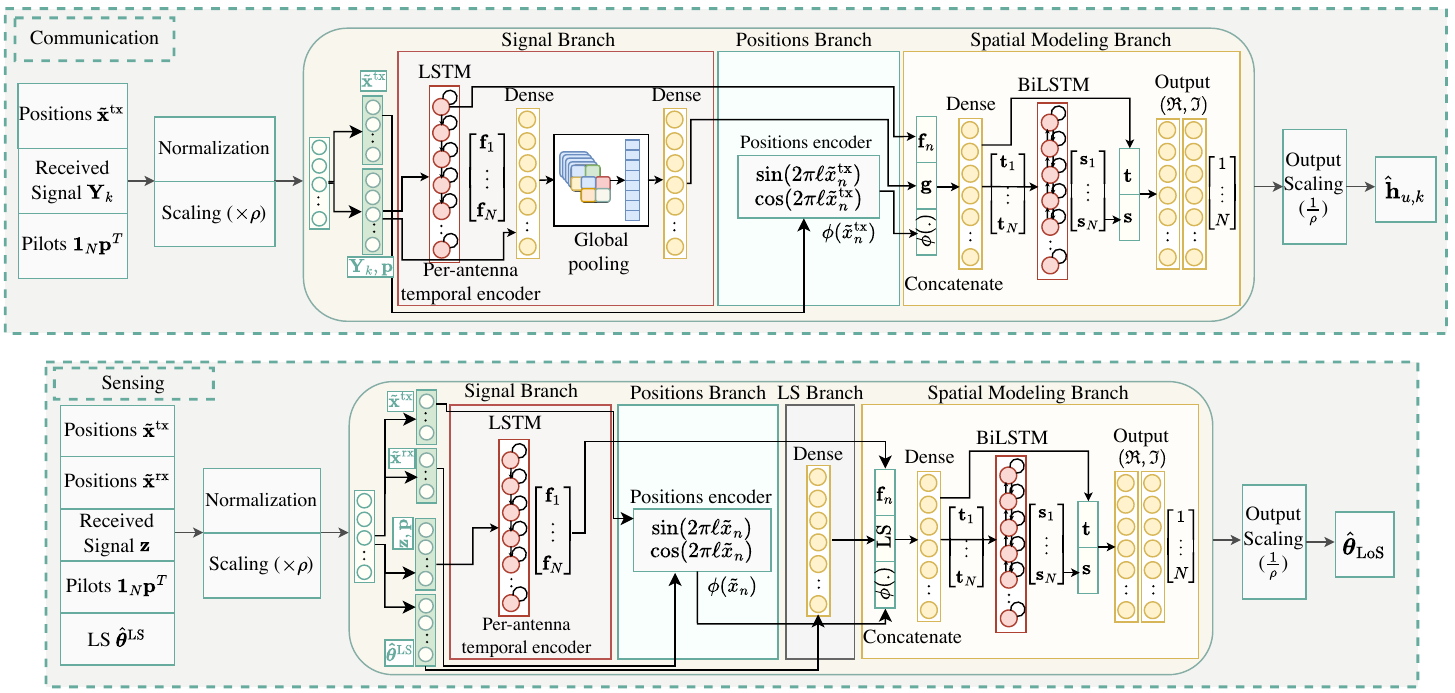}}
        \caption{Proposed LSTM-based channel estimation framework.}
        \label{Fig:lstm}
        \end{figure*}

		{To this end}, this work adopts an LSTM-based architecture that explicitly models the temporal evolution of pilot observations per antenna, while subsequently exploiting spatial correlations across antennas.
		{Particularly,} as illustrated in Fig. \ref{Fig:lstm}, the proposed estimator is organized into three main branches:
		(i) a temporal signal branch, where a shared LSTM processes the pilot-domain sequence of each PA to extract per-antenna temporal features;
		{(ii) a positional branch, which encodes the normalized PA locations using Fourier-based embeddings to capture both free-space geometry-dependent variations and in-waveguide phase progression; and}
		{(iii) a spatial modeling block, implemented using a bidirectional LSTM (BiLSTM) along the ordered PAs to capture inter-antenna dependencies induced by neighboring PA locations and the shared waveguide propagation path.}
		
		For the sensing link, an additional LS-based branch is incorporated to provide a coarse physics-driven initialization that is fused with the above branches.
		The combined features are finally passed through a lightweight per-antenna regression head, which maps the learned spatio-temporal representations to the real and imaginary components of the channel vector.


		\paragraph{Communication Link}
		
		A temporal sequence is first constructed for each transmit PA based on the input feature  $\mathbf{R}^{\mathrm{UE}_k}$ {in \eqref{eq:user_input}.}  
		For PA $n \in \{1,\ldots,N\}$, {define the real-valued sequence vector}
		$\mathbf{u}_n(\tau) \in \mathbb{R}^{2},  \tau = 1,\ldots,2T,$ as		
		\begin{equation}		
			\label{eq:temporal}
			\begin{split}			
				\mathbf{u}_n(\tau)
				&= \\
				& \begin{cases}
					\big[\Re\{[\mathbf{Y}_k]_{n,\tau}\},\,\Im\{[\mathbf{Y}_k]_{n,\tau}\}\big]^{\mathsf T}, & \tau=1,\dots,T,\\[4pt]
					\big[\Re\{p_{\tau-T}\},\, \Im\{p_{\tau-T}\}\big]^{\mathsf T}, & \tau=T+1,\dots,2T,
				\end{cases}
			\end{split}
		\end{equation}
		
		\noindent where $[\mathbf{Y}_k]_{n,\tau}$ denotes the $(n,\tau)$th entry of the lifted observation matrix $\mathbf{Y}_k$.  
		\textcolor{black}{Stacking these vectors yields $
		\mathbf{U}_n =
		[ \mathbf{u}_n(1),\,\ldots,\,\mathbf{u}_n(2T) ]^{\mathsf T}
		\in \mathbb{R}^{2T \times 2},$
		which preserves the temporal ordering of the received pilots and embeds the known pilot structure.} \textcolor{black}{This sequence is processed by a shared LSTM encoder with hidden dimension $
		H_u$. } 
		Let $\mathbf{h}_{n,\tau} \in \mathbb{R}^{H_u}$ and $\mathbf{c}_{n,\tau} \in \mathbb{R}^{H_u}$ denote the hidden and cell states at time step $\tau$. The gating variables consist of the input gate $\mathbf{i}_{n,\tau}$, forget gate $\mathbf{r}_{n,\tau}$, output gate $\mathbf{o}_{n,\tau}$, and candidate cell state $\tilde{\mathbf{c}}_{n,\tau}$, defined as
		\begin{align}
			\label{eq:lstm1}
			\mathbf{i}_{n,\tau} &= \sigma(\mathbf{W}_i \mathbf{u}_n(\tau) + \mathbf{U}_i \mathbf{h}_{n,\tau-1} + \mathbf{b}_i), \\
			\mathbf{r}_{n,\tau} &= \sigma(\mathbf{W}_r \mathbf{u}_n(\tau) + \mathbf{U}_r \mathbf{h}_{n,\tau-1} + \mathbf{b}_r), \\
			\mathbf{o}_{n,\tau} &= \sigma(\mathbf{W}_o \mathbf{u}_n(\tau) + \mathbf{U}_o \mathbf{h}_{n,\tau-1} + \mathbf{b}_o), \\
			\tilde{\mathbf{c}}_{n,\tau} &= \tanh(\mathbf{W}_c \mathbf{u}_n(\tau) + \mathbf{U}_c \mathbf{h}_{n,\tau-1} + \mathbf{b}_c),
		\end{align}
		\textcolor{black}{where $\mathbf{W}_\star\in\mathbb{R}^{H_u\times 2}$,  
		$\mathbf{U}_\star\in\mathbb{R}^{H_u\times H_u}$, and  
		$\mathbf{b}_\star\in\mathbb{R}^{H_u}$  
		denote the input weights, recurrent weights, and biases, respectively. $\sigma(\cdot)$ and $\tanh(\cdot)$ denote the sigmoid and hyperbolic tangent activations.} The cell and hidden states evolve as
		\begin{align}
			\label{eq:lstm2}
			\mathbf{c}_{n,\tau} &= \mathbf{r}_{n,\tau} \odot \mathbf{c}_{n,\tau-1}
			+ \mathbf{i}_{n,\tau} \odot \tilde{\mathbf{c}}_{n,\tau}, \\
			\mathbf{h}_{n,\tau} &= \mathbf{o}_{n,\tau} \odot \tanh(\mathbf{c}_{n,\tau}).
		\end{align}
		
		The temporal feature representation of the \textcolor{black}{$n$th PA}  is taken as the final hidden state
		$
		\mathbf{f}_n \triangleq \mathbf{h}_{n,2T} \in \mathbb{R}^{H_u}.
		$
		\textcolor{black}{Since all PAs share the same LSTM parameters, this encoder is applied identically across antennas.}

	The extracted temporal feature $\mathbf{f}_n$ captures the pilot-domain evolution associated with the $n$th PA. To incorporate geometry-dependent structure, the normalized PA position $\tilde{x}^{\mathrm{tx}}_n$ (i.e., from~\eqref{eq:user_input}) is mapped to a Fourier feature embedding
	\begin{equation}
		\label{eq:four}
		\boldsymbol{\phi}(\tilde{x}^{\mathrm{tx}}_n)
		=
		\big[
		\tilde{x}^{\mathrm{tx}}_n,\;
		\sin(2\pi \ell \tilde{x}^{\mathrm{tx}}_n),\;
		\cos(2\pi \ell \tilde{x}^{\mathrm{tx}}_n)
		\big]_{\ell=1}^{L_f},
	\end{equation}
	where $L_f$ denotes the number of Fourier modes. This embedding enables the NN to represent spatially periodic phase variations induced by waveguide propagation and array geometry.

	\textcolor{black}{In addition to the per-antenna temporal features,  a time-distributed dense layer, $\zeta(\cdot)$, with width $d_{z1}$ is applied to all antenna-time samples $\{\mathbf{u}_n(\tau)\}_{n,\tau}$. The resulting features are then aggregated via global average pooling to obtain the global context vector}
		\begin{equation}
		\mathbf{g}
		=
		\mathrm{Pool}\!\big(\zeta(\{\mathbf{u}_n(\tau)\}_{n,\tau})\big)
		\in \mathbb{R}^{d_z},
	\end{equation}
			\textcolor{black}{where $d_{z1}$ and $d_z$ denote the widths of the intermediate dense layer and the resulting context vector, respectively.} The resulting vector, $\mathbf{g}$, is broadcast to all antennas and concatenated with 
	$\mathbf{f}_n$ and ${\phi}(\tilde{x}^{\mathrm{tx}}_n)$, forming the geometry-aware token $
	\mathbf{t}_n^{(0)} 
	=
	\big[
	{\phi}(\tilde{x}^{\mathrm{tx}}_n),\;
	\mathbf{f}_n,\;
	\mathbf{g}
	\big].$ This token is passed through a linear projection with output width $d_{\mathrm{u}}$, followed by layer normalization as
	\begin{equation}
		\label{eq:token}
		\mathbf{t}_n
		=
		\mathrm{LN}\!\left(
		\mathbf{W}_p \mathbf{t}_n^{(0)} + \mathbf{b}_p
		\right)
		\in \mathbb{R}^{d_{\mathrm{u}}}.
	\end{equation}
	
	\noindent \textcolor{black}{Stacking the antenna-wise tokens yields $\mathbf{T}
	=
	\big[\mathbf{t}_1,\dots,\mathbf{t}_N\big]^{\mathsf T}
	\in
	\mathbb{R}^{N\times d_{\mathrm{u}}},$}
	which preserves the natural ordering of PAs along the waveguide.
	
\textcolor{black}{To explicitly model spatial dependencies across antennas, $\mathbf{T}$ is processed by a BiLSTM operating along the PAs. Let $\overrightarrow{\mathbf{h}}_n$ and $\overleftarrow{\mathbf{h}}_n$ denote the forward and backward hidden states, respectively. The resulting spatial representation at antenna $n$ is}
\begin{equation}
	\label{eq:bilstm}
\mathbf{s}_n
=
\big[
\overrightarrow{\mathbf{h}}_n,\;
\overleftarrow{\mathbf{h}}_n
\big]
\in
\mathbb{R}^{2H_s^{u}},
\end{equation}
\textcolor{black}{where $H_s^{u}$ denotes the hidden size per direction of the spatial BiLSTM. Stacking the antenna-wise representations yields}
$
\mathbf{S}
=
\big[\mathbf{s}_1,\dots,\mathbf{s}_N\big]^{\mathsf T}
\in
\mathbb{R}^{N\times 2H_s^{u}}.$
Through this recurrent processing, the network captures inter-antenna correlations induced by the PA arrangement and the deterministic waveguide propagation.

Finally, the spatial feature $\mathbf{s}_n$ is concatenated with the token $\mathbf{t}_n$ and passed through a two-layer feedforward regression head with \textcolor{black}{hidden width $d_h$} to predict the real and imaginary components of the channel. Collecting the per-antenna outputs yields
\begin{equation}
	\label{eq:commOutput}
	\hat{\mathbf{O}}^{\mathrm{UE}_k}
	=
	\big[
	\hat{\mathbf{h}}^{\Re}_{u,k},\;
	\hat{\mathbf{h}}^{\Im}_{u,k}
	\big]
	\in
	\mathbb{R}^{N\times 2},
\end{equation}
which corresponds to the estimated complex channel vector $\hat{\mathbf{h}}_{u,k}\in\mathbb{C}^{N\times 1}$.

\paragraph{Sensing Link}

The sensing estimator follows the same spatio-temporal design principle as the communication model, while incorporating additional structure through the richer input features defined in \eqref{eq:target_input}. In particular, beyond the lifted sensing observations and pilot symbols, the model explicitly includes (i) the normalized receive-PA position $\tilde{\mathbf{x}}^{\mathrm{rx}}$, and (ii) the switching-based LS estimate $\hat{\boldsymbol{\theta}}^{\mathrm{LS}}$ obtained in \eqref{eq:LS_target}.  Architecturally, the core building blocks remain unchanged: a weight-shared temporal LSTM extracts per-antenna sequential features, Fourier-based positional embeddings encode geometry-dependent structure, and a spatial BiLSTM models inter-antenna dependencies. The difference lies in how the additional sensing-specific features are embedded and fused with these components, as detailed next.

Concretely, for each PA, a temporal sequence is constructed from the sensing observation block and pilot symbols in the same manner as in the communication case. This per-antenna sequence is processed by a weight-shared temporal LSTM, producing a temporal feature vector $\mathbf{f}^{\mathrm{tar}}_n \in \mathbb{R}^{H_\mathrm{tar}}$, \textcolor{black}{where ${H_\mathrm{tar}}$ denotes the hidden dimension of the temporal LSTM.} In parallel, the normalized transmit and receive positions are encoded using Fourier embeddings, $\boldsymbol{\phi}_{\mathrm{tx}}(\tilde{x}^{\mathrm{tx}}_n)$ and $
\boldsymbol{\phi}_{\mathrm{rx}}(\tilde{x}^{\mathrm{rx}})$, which encode geometry-dependent phase variations along both waveguides.

Moreover, the LS estimate $\hat{\boldsymbol{\theta}}^{\mathrm{LS}}$ is embedded through a learnable linear transformation as
\begin{equation}
	\label{eq:lsemb}
\mathbf{e}^{\mathrm{LS}}_n
=
\mathbf{W}_{\mathrm{LS}}
\big[
\Re\{\hat{\boldsymbol{\theta}}^{\mathrm{LS}}(n)\},\;
\Im\{\hat{\boldsymbol{\theta}}^{\mathrm{LS}}(n)\}
\big]^{\mathsf T}
+
\mathbf{b}_{\mathrm{LS}}
\in
\mathbb{R}^{d_{\mathrm{LS}}},
\end{equation}
\textcolor{black}{thereby mapping the coarse estimate into a $d_{\mathrm{LS}}$-dimensional feature space suitable for nonlinear refinement.} The complete per-antenna sensing token is then formed as
\begin{equation}
\mathbf{t}^{\mathrm{tar}}_n
=
\mathrm{LN}\!\left(
\mathbf{W}_p
\big[
\boldsymbol{\phi}_{\mathrm{tx}}(\tilde{x}^{\mathrm{tx}}_n),\;
\boldsymbol{\phi}_{\mathrm{rx}}(\tilde{x}^{\mathrm{rx}}_n),\;
\mathbf{f}^{\mathrm{tar}}_n,\;
\mathbf{e}^{\mathrm{LS}}_n
\big]
+
\mathbf{b}_p
\right).
\end{equation}
Stacking the antenna-wise tokens yields $
\mathbf{T}^{\mathrm{tar}} =
[\mathbf{t}^{\mathrm{tar}}_1,\dots,\mathbf{t}^{\mathrm{tar}}_N]^{\mathsf T},$ which is then processed by the spatial BiLSTM, similar to the communication model. This spatial block captures inter-antenna correlations induced by the transmit waveguide, while implicitly accounting for the single receive-PA coupling embedded in the features. Finally, the spatial representation $\mathbf{s}^{\mathrm{tar}}_n$ is concatenated with the token $\mathbf{t}^{\mathrm{tar}}_n$, and then passed through a feedforward regression head to produce the real and imaginary components of the LoS sensing channel,
yielding the estimate
\begin{equation}
	\label{eq:sensingOutput}
\hat{\mathbf{O}}^{\mathrm{BS}}
=
\big[
\hat{\boldsymbol{\Theta}}^{\mathrm{Re}}_{\mathrm{LoS}},\;
\hat{\boldsymbol{\Theta}}^{\mathrm{Im}}_{\mathrm{LoS}}
\big]
\in
\mathbb{R}^{N\times 2}.
\end{equation}

\noindent \textcolor{black}{Fig.~\ref{Fig:lstm} provides an overview of the proposed framework and illustrates the flow of information through both the SAC estimation stages.}

		\paragraph{Training and Testing}
	
		The NN parameters are optimized using the mean absolute error between the predicted and true channels. Specifically, the loss function is defined as
		\begin{equation}
			\label{eq:loss}
		\mathcal{L}_{\mathrm{MAE}}
		=
		\mathbb{E}\!\left[
		\left\|
		\hat{\mathbf{O}}^{o}
		-
		{\mathbf{O}}^{o}
		\right\|_{1}
		\right],
	\end{equation}
		where  $o\in \{{\mathrm{BS}}, \mathrm{UE}_k\}$ and $\|\cdot\|_{1}$ denotes the element-wise $\ell_1$-norm applied to the real-imaginary channel representation.

		{The SAC datasets are formed by pairing the input feature representations with their corresponding channel labels according to \eqref{eq:user_input}--\eqref{eq:outputBS}.} Prior to network training, both the SAC datasets undergo normalization in order to improve numerical stability and accelerate convergence. The same preprocessing steps are applied to both SAC tasks for consistency.

		Let 
        {$\mathbf{X}^{o}$ denote any real-valued input feature extracted from  $\mathbf{R}^{\mathrm{BS}}$ or $\mathbf{R}^{\mathrm{UE}_k}$ (e.g., normalized PA positions).} The standardized feature is computed as
		\begin{equation}
			\label{eq:std}
			\mathbf{X}_{\mathrm{std}}^{o}(b,v)
			=
			\frac{
				\mathbf{X}^{o}(b,v)
				-
				\mathbb{E}\!\left\{\mathbf{X}^{o}(b,v)\right\}
			}{
				\mathbb{D}\!\left\{\mathbf{X}^{o}(b,v)\right\}
			},
		\end{equation}
		\noindent where $\mathbb{E}\{\cdot\}$ and $\mathbb{D}\{\cdot\}$ denote the empirical mean and standard deviation computed over the training {dataset}. Here, $b \in \{1,\dots,B\}$ denotes the training sample index, {where $B$ is the total number of training samples}, and $v$ \textcolor{black}{is the index corresponding to the feature dimension}. For the communication task, \eqref{eq:std} is applied to the transmit-PA position feature.  
		For the sensing task, both the transmit- and receive-PA position features are standardized according to \eqref{eq:std}, using their respective training-set statistics.

		Furthermore, during dataset generation, the {received complex} pilot signals  are multiplied by a constant scaling factor, $\rho$, to control the dynamic range of the input features (i.e., 
			$\rho \mathbf{Y}_k, \rho \mathbf{z}$). Additionally, the complex channel labels are normalized by a global root-mean-square (RMS) factor computed from the training {dataset}, defined as

		\begin{equation}
		\rho^{\mathrm{o}}
		=
		\left(
		\mathbb{E}\!\left[
		\left\|
		\mathbf{O}^{\mathrm{o}}(b,n)
		\right\|_2^2
		\right]
		\right)^{-\frac{1}{2}},
	\end{equation}

		\noindent and the scaled label becomes
		\begin{equation}
			\label{eq:scal}
			\mathbf{O}_{\mathrm{scaled}}^{\mathrm{o}}(b,n)
			=
			\rho^{\mathrm{o}}\,
			\mathbf{O}^{\mathrm{o}}(b,n).
		\end{equation}
		
		\noindent These preprocessing steps stabilize gradient magnitudes and improve convergence for both estimation tasks. {The overall procedure of the proposed LSTM-based channel estimation framework is summarized in Algorithm~\ref{alg:lstm_estimator}.}
		

			\newcommand{\LongState}[1]{%
	\State \begin{minipage}[t]{0.9\linewidth}
		#1
	\end{minipage}
}

\begin{algorithm}[t]
	\caption{Proposed LSTM-Based Channel Estimation Framework}
	\label{alg:lstm_estimator}
	{
		\begin{algorithmic}[1]
			\Require $(\mathbf R_{\mathrm{UE}_k},\mathbf O_{\mathrm{UE}_k})$, $(\mathbf R_{\mathrm{BS}},\mathbf O_{\mathrm{BS}})$;
			\Ensure Trained communication and sensing estimators;
			
			\Statex \textbf{Offline Training:}
			
%
%
			
			\LongState{Normalize the input features and scale the channel labels according to \eqref{eq:std}--\eqref{eq:scal};}
			
			
			\While{$i \leq E_\epsilon$ or the validation accuracy increases over $15$ epochs}
			
			\State{Construct the per-PA temporal sequence according to \eqref{eq:temporal};}
			\State{Extract temporal features using \eqref{eq:lstm1}--\eqref{eq:lstm2}}
			\State{Encode PA positions using \eqref{eq:four};}
			\State{For sensing, embed the LS-assisted feature using \eqref{eq:lsemb};}
			\LongState{Generate the PA-wise feature representation using \eqref{eq:token};}
			\State{Apply the spatial BiLSTM according to \eqref{eq:bilstm};}
			\State{Obtain the channel estimates using \eqref{eq:commOutput} or \eqref{eq:sensingOutput};}

			\LongState{Update the network by minimizing $\mathcal{L}_{\mathrm{MAE}}$ in \eqref{eq:loss};}
			
			\EndWhile
			
			
			\Statex \textbf{Online Testing:}
			
			\LongState{Given new received observations, $\mathbf R_{\mathrm{UE}_k}^{\rm test}$ and $\mathbf R_{\mathrm{BS}}^{\rm test}$;}
			
%
			
			\State Estimate $\hat{\mathbf O}_{\mathrm{UE}_k}$ using the trained UE estimator;
			\State Estimate $\hat{\mathbf O}_{\mathrm{BS}}$ using the trained BS estimator;
			
			\State \Return $\hat{\mathbf h}_{u,k}$ and $\hat{\boldsymbol{\Theta}}_{\mathrm{LoS}}$.
	\end{algorithmic}}
\end{algorithm}

	\subsection{Sensing Parameter Estimation {Framework}} 
	\label{Sec:Music}

\begin{algorithm}[t]
	\caption{2D Near-Field MUSIC}
	\label{alg:music}
	{
		\begin{algorithmic}[1]
			\Require $\{\hat{\boldsymbol{\Theta}}_{\mathrm{LoS}}[m]\}_{m=1}^{M}$,
			$\{\boldsymbol{\psi}^{\mathrm{tx}}_{n}\}_{n=1}^{N}$,
			$\boldsymbol{\psi}^{\mathrm{rx}}$, and $\mathcal{G}$;
			\Ensure Estimated target range and azimuth angle $(\hat r,\hat\varphi)$;
			
			\State Construct the sample covariance matrix as in \eqref{eq:sample_cov};
			
			\LongState{Perform eigenvalue decomposition of $\hat{\mathbf{R}}_{\theta}$:
				$
				\hat{\mathbf{R}}_{\theta}
				=
				\mathbf{U}\boldsymbol{\Lambda}\mathbf{U}^{H};
				$}
			
			\LongState{Sort the eigenvalues in descending order and form the noise-subspace matrix
				$\mathbf{E}_{n}\in\mathbb{C}^{N\times(N-1)}$ using the eigenvectors associated with the smallest $N-1$ eigenvalues;}
			
			\For{each $r\in[r_{\min},r_{\max}]$}
			\For{each $\varphi\in[\varphi_{\min},\varphi_{\max}]$}
			
			\LongState{Compute $\boldsymbol{\psi}_{\mathrm{tar}}(r,\varphi)$,
				$d^{\mathrm{tx}}_{n}(r,\varphi)$, and
				$d^{\mathrm{rx}}(r,\varphi)$ according to
				\eqref{eq:d_tx_music} and \eqref{eq:d_rx_music};}
			
			\LongState{Construct the near-field steering vector
				$\mathbf{a}(r,\varphi)$ according to \eqref{eq:nf_steering};}
			
			\LongState{Normalize the steering vector as
				\[
				\bar{\mathbf{a}}(r,\varphi)
				=
				\frac{\mathbf{a}(r,\varphi)}
				{\|\mathbf{a}(r,\varphi)\|_{2}};
				\]}
			
			\LongState{Evaluate the 2D near-field MUSIC pseudospectrum as in \eqref{eq:music_spec};}
			
			\EndFor
			\EndFor
			
			\State Locate the dominant peak of the MUSIC spectrum as in \eqref{eq:musiclast};
			
			\State \Return $(\hat r,\hat\varphi)$.
	\end{algorithmic}}
\end{algorithm}
		
		After obtaining the estimated LoS sensing channel associated with the target,
		$\hat{\boldsymbol{\theta}}_{\mathrm{LoS}} \in \mathbb{C}^{N\times 1}$,
		we next extract the target  parameters, i.e., its azimuth angle and range, by applying a 2D near-field  {MUSIC algorithm}. {It is important to emphasize that, unlike conventional ISAC systems where the received antenna-array snapshots can be directly used to form a spatial covariance matrix \cite{MUSIC, Faisal2026HybridAngleParametric}, MUSIC cannot be applied directly to the received pilot observations in PA-assisted ISAC systems. Due to the waveguide combining, each pilot symbol yields only a single scalar measurement, which provides an insufficient spatial aperture to form a meaningful covariance matrix. By first estimating the LoS sensing channel $\hat{\boldsymbol{\theta}}_{\mathrm{LoS}}$, the proposed framework reconstructs an $N$-dimensional spatial signature that enables reliable subspace-based parameter extraction.}
		Unlike far-field {MUSIC}, where the array response is modeled via plane waves, here a spherical-wave model is adopted, which accounts for the distance-dependent phase and amplitude variations across the PAs. Let the target be parameterized in polar form on the $x$-$y$ plane as $
			\boldsymbol{\psi}_{\mathrm{tar}}(r,\varphi)=
			\big(r\cos\varphi,\; r\sin\varphi,\; 0\big),$
		where $r$ and $\varphi$ denote the range and azimuth angle with respect to the origin of the considered coordinate system. Define the bistatic distances {as}
		
		\begin{align}
			d^{\mathrm{tx}}_n(r,\varphi) &\triangleq \big\| \boldsymbol{\psi}^{\mathrm{tx}}_n - \boldsymbol{\psi}_{\mathrm{tar}}(r,\varphi)\big\|, \label{eq:d_tx_music}\\
			d^{\mathrm{rx}}(r,\varphi) &\triangleq \big\| \boldsymbol{\psi}^{\mathrm{rx}} - \boldsymbol{\psi}_{\mathrm{tar}}(r,\varphi)\big\|. \label{eq:d_rx_music}
		\end{align}
		Under the spherical-wave propagation model used in Section~\ref{Sec:System_Model}, the LoS sensing channel can be written (up to a scalar factor)
		as a deterministic function of $(r,\varphi)$ \textcolor{black}{according to}
		\begin{equation}\label{eq:theta_param}
			\boldsymbol{\Theta}_{\mathrm{LoS}}(r,\varphi)
			=
			\gamma\;
			\mathbf{a}(r,\varphi),
		\end{equation}
		where $\gamma \in \mathbb{C}$ {accounts for} the unknown complex reflectivity and common propagation constants, and
		$\mathbf{a}(r,\varphi)\in\mathbb{C}^{N\times 1}$ is the near-field steering vector with {its $n$th entry as}
		\begin{equation}\label{eq:nf_steering}
			\begin{split}
			\big[\mathbf{a}(r,\varphi)\big]_n
			=  \frac{\exp\!\left(-j\frac{2\pi}{\lambda}\big(d^{\mathrm{rx}}(r,\varphi)+d^{\mathrm{tx}}_n(r,\varphi)\big)\right)}{{d^{\mathrm{rx}}(r,\varphi)\, d^{\mathrm{tx}}_n(r,\varphi)}}.
			\end{split}
		\end{equation}

		\noindent Note that \eqref{eq:nf_steering} generalizes the conventional far-field array manifold by explicitly incorporating
		(i) the spherical phase term determined by the bistatic path length, and (ii) the distance-dependent attenuation across antennas. Since MUSIC is invariant to an unknown complex scaling, the factor $\gamma$ does not affect parameter recovery. MUSIC is applied using the
		normalized steering vector $\bar{\mathbf a}(r,\varphi)=\mathbf a(r,\varphi)/\|\mathbf a(r,\varphi)\|_2$.
		

		To apply subspace-based estimation, {${M}$ snapshots} of the estimated LoS sensing channel are collected
		$\{\hat{\boldsymbol{\Theta}}_{\mathrm{LoS}}[m]\}_{m=1}^{{M}}$ and the sample covariance matrix is then constructed as
		\begin{equation}\label{eq:sample_cov}
			\hat{\mathbf{R}}_{\theta}
			=
			\frac{1}{{M}}\sum_{m=1}^{{M}} \hat{\boldsymbol{\Theta}}_{\mathrm{LoS}}[m]\,
			\hat{\boldsymbol{\Theta}}_{\mathrm{LoS}}^{\mathsf H}[m]
			\in \mathbb{C}^{N\times N}.
		\end{equation}

			
		\noindent Performing eigenvalue decomposition of $\hat{\mathbf R}_{\theta}$ yields the noise-subspace basis
		$\mathbf E_{\mathrm n}\in\mathbb{C}^{N\times(N-1)}$, formed by the eigenvectors
		associated with the smallest $N-1$ eigenvalues.
		
		
		To this end, the 2D near-field MUSIC pseudospectrum is evaluated over a predefined search grid
		$\mathcal{G}=\{(r,\varphi)$: $r\in[r_{\min},r_{\max}],\;\varphi\in[\varphi_{\min},\varphi_{\max}]\}$ as
		\begin{equation}\label{eq:music_spec}
			P_{\mathrm{MUSIC}}(r,\varphi)
			=
			\frac{1}{\bar{\mathbf a}^{\mathsf H}(r,\varphi)\,\mathbf{E}_{\mathrm{n}}\mathbf{E}_{\mathrm{n}}^{\mathsf H}\,\bar{\mathbf a}(r,\varphi)}.
		\end{equation}

		\noindent Finally, the target parameters are then obtained by locating the dominant peak of \eqref{eq:music_spec} as
		\begin{equation} \label{eq:musiclast}
			(\hat{r},\hat{\varphi})
			=
			\arg\max_{(r,\varphi)\in\mathcal{G}}
			P_{\mathrm{MUSIC}}(r,\varphi).
		\end{equation}
		
		\noindent {In the considered discrete PA activation model, the MUSIC pseudospectrum stability is influenced by the effective aperture and spatial distribution of the activated PAs. Well-distributed PA locations can reduce the correlation among nearby range-angle steering vectors and produce more distinguishable MUSIC peaks, whereas clustered locations may lead to a less stable pseudospectrum. The steps of the 2D near-field MUSIC procedure are summarized in Algorithm~\ref{alg:music}.}

\section{Complexity Analysis}\label{Sec:Complexity_Analysis}

This section analyzes the computational complexity of the proposed LSTM-based estimators in terms of the number of \textcolor{black}{real-valued multiplications} $C_M$ and additions $C_A$ required \textcolor{black}{for inference}. Both the UE and sensing-channel estimators share the same computational backbone, consisting of three main blocks: (i) a per-PA temporal LSTM encoder applied over the pilot dimension, (ii) a spatial BiLSTM applied over the ordered PAs, and (iii) a feed-forward dense  projection. Therefore, the total complexity is obtained by summing the costs of these blocks, with minor differences in the input feature dimensions. Consider a uni-directional LSTM with input dimension $D$ and hidden dimension $H$, processed over a sequence of length $L$. {Using the standard four-gate formulation \cite{6789445}, the number of real multiplications and additions per time step is given by}
\begin{align}
	C_M^{\mathrm{LSTM}}(D,H) &= 4(D+H)H + 3H, \label{eq:lstm_cm}\\
	C_A^{\mathrm{LSTM}}(D,H) &= 4(D+H)H + 5H. \label{eq:lstm_ca}
\end{align}
Hence, the total cost over $L$ steps is $L\,C_M^{\mathrm{LSTM}}(D,H)$ multiplications and $L\,C_A^{\mathrm{LSTM}}(D,H)$ additions. A BiLSTM with the same $(D,H)$ processes the sequence in two directions; thus, its cost is defined as
\begin{align}
	C_M^{\mathrm{BiLSTM}}(D,H) &= 2\,C_M^{\mathrm{LSTM}}(D,H), \label{eq:bilstm_cm}\\
	C_A^{\mathrm{BiLSTM}}(D,H) &= 2\,C_A^{\mathrm{LSTM}}(D,H). \label{eq:bilstm_ca}
\end{align}

\noindent A dense layer mapping $\mathbb{R}^{D_{\mathrm{in}}}\!\to\!\mathbb{R}^{D_{\mathrm{out}}}$ requires
\begin{equation}\label{eq:dense_cost}
	C_M^{\mathrm{FFD}}(D_{\mathrm{in}},D_{\mathrm{out}})
	=
	C_A^{\mathrm{FFD}}(D_{\mathrm{in}},D_{\mathrm{out}})
	=
	D_{\mathrm{in}}D_{\mathrm{out}}.
\end{equation}

\subsection{\textcolor{black}{Communication Channel Estimator Complexity}}

For the communication estimator, the per-PA temporal input at each time step has dimension $\mathbf{u}_n(\tau)\in\mathbb{R}^{2}$ and sequence length $2T$, as described in Sec.~\ref{sec:DLEst}. Accordingly, the temporal LSTM complexity is
\begin{align}
	C_{M,\mathrm{u}}^{\mathrm{temp}} &= 2N\,T\,C_M^{\mathrm{LSTM}}(2,H_u), \\
	C_{A,\mathrm{u}}^{\mathrm{temp}} &= 2N\,T\,C_A^{\mathrm{LSTM}}(2,H_u).
\end{align}
Next, the positional Fourier embedding of the transmit-PA locations contributes with  $C_{M,\mathrm{u}}^{\mathrm{Fourier}} = 2N L_f$.  The global pilot-context module consists of a time-distributed dense layer followed by global average pooling and a projection to the context vector. Its complexity is expressed as
\begin{align}
	C_{M,\mathrm{u}}^{z}
	&=
	2NT\, C_M^{\mathrm{FFD}}(2,d_{z1})
	+
	d_{z1}
	+
	C_M^{\mathrm{FFD}}(d_{z1},d_{z}), \\
	C_{A,\mathrm{u}}^{z}
	&=
	2NT\, C_A^{\mathrm{FFD}}(2,d_{z1})
	+
	d_{z1}(2NT\hspace{-1mm}-\hspace{-1mm}1)
	+
	C_A^{\mathrm{FFD}}(d_{z1},d_{z}).
\end{align}
Furthermore, the token projection maps the concatenated feature vector to width $d_u$.  The input dimension of this projection is $
	D_{\mathrm{u}}^{\mathrm{tok}} = (1+2L_f) + H_u + d_z$. {Thus, it contributes with}
\begin{align}
	C_{M,\mathrm{u}}^{\mathrm{tok}} &= N\,C_M^{\mathrm{FFD}}(D_{\mathrm{u}}^{\mathrm{tok}},d_u), \\
	C_{A,\mathrm{u}}^{\mathrm{tok}} &= N\,C_A^{\mathrm{FFD}}(D_{\mathrm{u}}^{\mathrm{tok}},d_u).
\end{align}
The spatial BiLSTM operating along PAs contributes with
\begin{align}
	C_{M,\mathrm{u}}^{\mathrm{spat}} &= N\,C_M^{\mathrm{BiLSTM}}(d_u,H_u^{\mathrm{s}}), \\
	C_{A,\mathrm{u}}^{\mathrm{spat}} &= N\,C_A^{\mathrm{BiLSTM}}(d_u,H_u^{\mathrm{s}}).
\end{align}
Finally, the two-layer regression contributes with
\begin{align}
	C_{M,\mathrm{u}}^{\mathrm{head}}
	&=
	N\!\left(
	C_M^{\mathrm{FFD}}(d_u+2H_u^{\mathrm{s}},d_h)
	+
	C_M^{\mathrm{FFD}}(d_h,2)
	\right), \\
	C_{A,\mathrm{u}}^{\mathrm{head}}
	&=
	N\!\left(
	C_A^{\mathrm{FFD}}(d_u+2H_u^{\mathrm{s}},d_h)
	+
	C_A^{\mathrm{FFD}}(d_h,2)
	\right).
\end{align}

Therefore, the total complexity is
\begin{align}
	C_{M}^{\mathrm{UE}} &= C_{M,\mathrm{u}}^{\mathrm{temp}} + C_{M,\mathrm{u}}^{z} + C_{M,\mathrm{u}}^{\mathrm{Fourier}} + C_{M,\mathrm{u}}^{\mathrm{tok}} + C_{M,\mathrm{u}}^{\mathrm{spat}} + C_{M,\mathrm{u}}^{\mathrm{head}}, \label{eq:cm_ue}\\
	C_{A}^{\mathrm{UE}} &= C_{A,\mathrm{u}}^{\mathrm{temp}} + C_{A,\mathrm{u}}^{z} + C_{A,\mathrm{u}}^{\mathrm{tok}} + C_{A,\mathrm{u}}^{\mathrm{spat}} + C_{A,\mathrm{u}}^{\mathrm{head}}. \label{eq:ca_ue}
\end{align}
{For $K$ UEs, the communication estimator is independently applied to each UE with shared network parameters, yielding linear inference complexity scaling with $K$.}
\subsection{\textcolor{black}{Sensing Channel Estimator Complexity}}

The sensing estimator follows the same computational backbone and thus admits the same block-wise structure as 
\eqref{eq:cm_ue}--\eqref{eq:ca_ue}, with the corresponding sensing parameters 
$(H_{\mathrm{tar}}, d_{\mathrm{tar}}, H_{\mathrm{tar}}^{\mathrm{s}})$ replacing 
$(H_{\mathrm{u}}, d_{\mathrm{u}}, H_{\mathrm{u}}^{\mathrm{s}})$. 
In addition, the sensing architecture includes positional Fourier embeddings for both transmit and receive positions and an LS-feature embedding implemented by a dense projection. The positional Fourier embedding of the transmit and receive locations contributes with $
	C_{M,\mathrm{tar}}^{\mathrm{Fourier}} = 4N L_f,$ following the same counting convention as above. The LS-feature embedding contributes with
\begin{align}
	C_{M,\mathrm{tar}}^{\mathrm{LS}} &= N\,C_M^{\mathrm{FFD}}(2,d_{\mathrm{LS}}),\\
	C_{A,\mathrm{tar}}^{\mathrm{LS}} &= N\,C_A^{\mathrm{FFD}}(2,d_{\mathrm{LS}}).
\end{align}
Therefore, the total complexity is expressed as
\begin{align}
	C_{M}^{\mathrm{BS}} &= C_{M,\mathrm{tar}}^{\mathrm{temp}} \hspace{-1mm}  +  C_{M,\mathrm{tar}}^{\mathrm{Fourier}} \hspace{-1mm} + C_{M,\mathrm{tar}}^{\mathrm{LS}} \hspace{-1mm} + C_{M,\mathrm{tar}}^{\mathrm{tok}} \hspace{-1mm} + C_{M,\mathrm{tar}}^{\mathrm{spat}} \hspace{-1mm} +  C_{M,\mathrm{tar}}^{\mathrm{head}}, \label{eq:cm_bs}\\
	C_{A}^{\mathrm{BS}} &= C_{A,\mathrm{tar}}^{\mathrm{temp}} + C_{A,\mathrm{tar}}^{\mathrm{LS}} + C_{A,\mathrm{tar}}^{\mathrm{tok}} + C_{A,\mathrm{tar}}^{\mathrm{spat}} + C_{A,\mathrm{tar}}^{\mathrm{head}}. \label{eq:ca_bs}
\end{align}

\noindent Overall, both estimators scale linearly with the number of PAs $N$ and the pilot-sequence length, while the dominant terms arise from the temporal LSTM and spatial BiLSTM blocks. To complement the complexity characterization, Table~\ref{tab:network_architecture} summarizes the  NN architectures of the proposed SAC estimators.

\section{Simulation Results}~\label{Sec:Simulation_Results}
This section comprehensively evaluates the performance of the proposed \textcolor{black}{PA-enabled} ISAC channel estimation framework in terms of estimation accuracy, sensing capability, and computational complexity under various system configurations {(i.e., SNR, $T$, and $N$)}. 
	\subsection{Simulation Parameters}
	\def\leftvspace{\rule{0pt}{4.5mm}}
\def\rightvspace{2mm}

\begin{table}[h!]
	\centering
	\caption{Proposed Estimators Architecture.}
	\label{tab:network_architecture}
	\begin{tabular}{c|l|c|c}
		\textbf{Model} &	\textbf{Layers} & \textbf{Size}  & \textbf{Activation} \\  \hline
		\multirow{9}{*}{\rotatebox{90}{Communication}}  &  Input  & $N\times(4T+1)$  & - \\
		&  Temporal LSTM    &  $H_{\mathrm{u}} = 48$      &  $\tanh/\sigma$ \\
		&  Time-distributed dense    & $d_{z1} = 128$     & $\mathrm{Gelu}$ \\
		&  Global context dense    & $d_{z} = 64$      & $\mathrm{Gelu}$ \\
		&  Positional embedding  &  $L_f = 4$       & {-} \\
		&  Projection dense   & $d_{\mathrm{u}} = 128$      & $\mathrm{Gelu}$ \\
		&  Spatial BiLSTM    & $H_{\mathrm{u}}^{\mathrm{s}} = 32$     & $\tanh/\sigma$ \\
		&  Head dense &  $d_{\mathrm{h}} = 64$      & $\mathrm{Gelu}$ \\
		&  Output & $N \times 2$    & - \\  \hline 

		\multirow{8}{*}{\rotatebox{90}{Sensing}}  & Input  & $N \times (4T + 4)$   & - \\
		&  Temporal LSTM    &  $H_{\mathrm{tar}} = 64$      &  $\tanh/\sigma$ \\
		& LS embedding & $d_{\mathrm{LS}} = 32$ & $\mathrm{Gelu}$  \\
		&  Positional embedding  &  $L_f = 4$       & {-} \\
		&  Projection dense   & $d_{\mathrm{tar}} = 96$      & $\mathrm{Gelu}$ \\
		&  Spatial BiLSTM    & $H_{\mathrm{tar}}^{\mathrm{s}} = 32$    & $\tanh/\sigma$ \\
		&  Head dense  &  $d_{\mathrm{h}} = 64$      & $\mathrm{Gelu}$ \\
		&  Output & $N \times 2$    & - \\  \hline 

	\end{tabular}
\end{table}

	{The simulation results presented in this section assume} $K = 3$, $Q=200$, $N = 16$, and $T=8$  unless otherwise stated. A 3D Cartesian coordinate system is considered, where the transmit dielectric waveguide is placed parallel to the $x$-axis at a height of $d = 5$~m \cite{PA_channel}. The coverage region lies on the $x$--$y$ plane with dimensions $D_x \times D_y = 20 \times 20$~m$^2$. The carrier frequency is set to $f_c = 28$~GHz, yielding a wavelength $\lambda = c/f_c$, where $c = 3 \times 10^8$~m/s denotes the speed of light. {The effective refractive index of the dielectric waveguide is $\eta_{\mathrm{eff}} = 1.4$ \cite{PA_channel, PA_channel2}. This parameter determines the guided wavelength and the resulting phase progression across the PAs, which is incorporated into the position-aware feature representation, enabling the proposed estimator to accommodate variations in the waveguide-induced propagation characteristics.} The candidate PA locations along the waveguide are uniformly spaced with a minimum spacing of $\delta = 0.1$~m. For each realization, $N$ PAs are randomly selected from the pre-defined discrete positions. The transmit waveguide feed point is located at $\boldsymbol{\psi}_0 = (0,0,d)$, whereas the receive waveguide feed point is located at $\boldsymbol{\psi}^{\mathrm{rx}}_0 = (0,y^{\mathrm{rx}},d)$ with $y^{\mathrm{rx}}=5$~m. 
	\textcolor{black}{The target has the polar coordinates $(r,\varphi)$, where $r=10$~m and $\varphi=\pi/9$, corresponding to the Cartesian position
	$\boldsymbol{\psi}_{\mathrm{tar}}=(r\cos\varphi,\; r\sin\varphi,\;0)$.} The target reflection coefficient is modeled as a unit-modulus complex scalar with random phase uniformly distributed over $[0,2\pi)$. 
	The propagation environment includes \textcolor{black}{three scattered components (i.e., $S=3$)}, {whose spatial coordinates are independently sampled from uniform distributions within the considered region.} {This uniform distribution provides a simplified representation of multipath-rich environments without modeling specific urban geometries or material properties.}

	The training dataset size is  configured to {contain $B=10^5$ samples per SNR level for both SAC tasks.} For each configuration, $90\%$ of the samples are used for training and the remaining $10\%$ for validation. During the training phase, the SNR is fixed at $\mathrm{SNR}=20$~dB. In contrast, during the testing phase, a wider SNR range of $\text{SNR} = \unit[-20:5:20]{dB}$ is considered. {For each SNR value, the reported NMSE results are averaged over $500$ independent unseen test samples.} {Here, the SNR is defined as the ratio between the received signal power and the AWGN power.}  
    This setup ensures that the proposed model is evaluated not only on unseen channel realizations but also under noise conditions outside its training regime. To assess generalization with respect to the array size, the model is trained using $N^{\mathrm{tr}} = 16$ PAs, while testing is performed over a broader set $N^{\mathrm{te}} = 8:4:32$. This strategy demonstrates the robustness {against system reconfiguration} and scalability of the proposed estimator, eliminating the need for {retraining or explicit SNR estimation at the receiver in practical deployments.} The model is trained using the Adam optimizer with a learning rate of \(2 \times 10^{-4}\). To reduce correlation among training examples, minibatches of $256$ randomly sampled {instances are employed.} Training is also governed by an early-stopping condition: it terminates if the {validation loss does not improve} for $15$ consecutive epochs or when the maximum epoch limit of \(E_\epsilon = 1000\) is reached. {To ensure reproducibility, a fixed random seed of zero is used for the training procedure.}

	
	For the MUSIC-based sensing evaluation, at each SNR value, ${M} = 50$ snapshots of the estimated LoS sensing channel are generated with a fixed PA geometry, and the sample covariance $\hat{\mathbf R}_\theta$ in \eqref{eq:sample_cov} is formed from these snapshots. For every SNR point, \textcolor{black}{$10$ independent} Monte Carlo trials are carried out, where the PA positions and small-scale fading are re-drawn between trials. The 2D MUSIC pseudospectrum in \eqref{eq:music_spec} is evaluated over a range grid $r \in [0,20]~\mathrm{m}$ with step size $0.05~\mathrm{m}$ and an angle grid $\varphi \in [0^\circ,180^\circ]$ with resolution $0.1^\circ$, and the resulting \textcolor{black}{angle of arrival (AoA)} and range root mean square errors (RMSEs) are obtained by averaging over all Monte Carlo trials at each SNR.


	 {To assess the performance of the proposed approach, the normalized mean square error (NMSE) is utilized as the primary evaluation metric \cite{Yu_ELM}. For the communication channel estimation task, the scheme proposed in~\cite{PA_channel}, namely PAformer, is adopted in addition to the switching-based LS (i.e., presented in Section~\ref{sec:LS}) and LMMSE estimators as benchmarks, providing a meaningful comparison with existing PA-based communication systems and classical estimators.  For the sensing channel estimation, to the best of our knowledge, this is the first work addressing channel estimation in PA-assisted ISAC systems. No prior learning-based or model-driven benchmark exists for this scenario. Therefore, the switching-based LS and LMMSE estimators are employed as classical baselines. These baselines provide references to evaluate the effectiveness and robustness of the proposed deep learning framework.}


	\subsection{Impact of Varying {the} SNR}


	\begin{figure}[!t]
	\centering
	\subfloat[]{
		\includegraphics[width=0.49\textwidth]{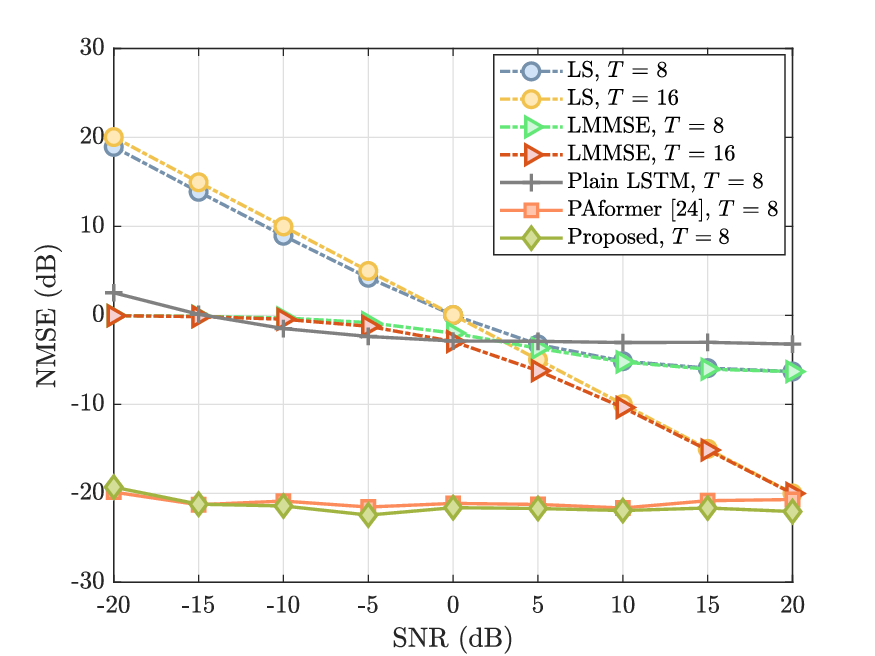}
		\label{Fig:Vary_SNR_u}
	}
	\subfloat[]{
		\includegraphics[width=0.49\textwidth]{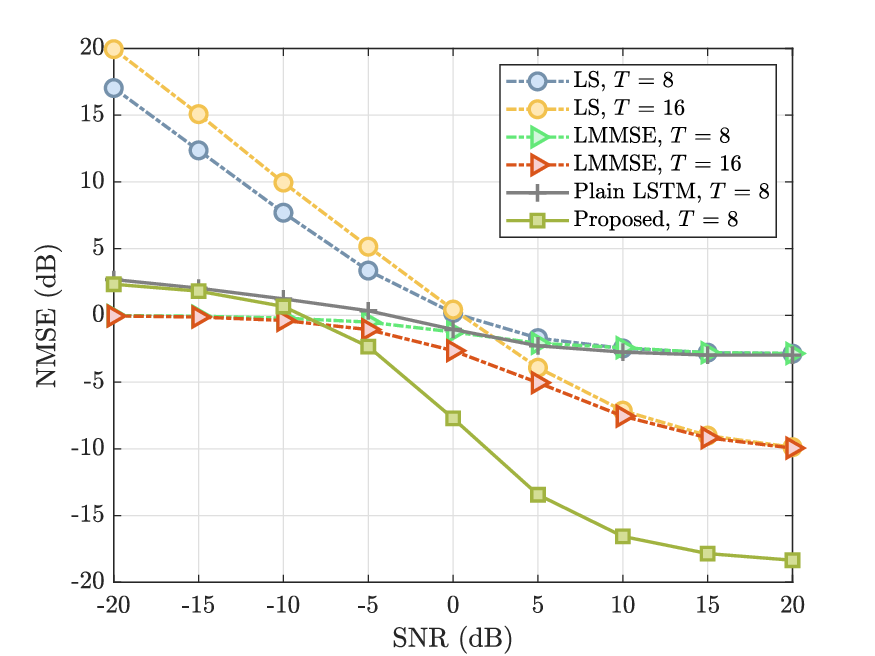}
		\label{Fig:Vary_SNR_t}
	}
	\caption{{NMSE performance versus SNR at $N = 16$. (a) Communication, (b) Sensing. }}
	\label{Fig:VarySNR}
\end{figure}

%


   {Fig.~\ref{Fig:VarySNR} illustrates the performance of NMSE versus SNR with varying value of $T\in \{8, 16\}$ and $N=16$ activated PAs {for both SAC links.}}
    Specifically, Fig.~\ref{Fig:Vary_SNR_u} depicts the performance of estimating the communication channel, $\mathbf{h}_{u,k}$. {It can be observed that both LS and LMMSE exhibit pronounced performance degradation when {$T<N$}, resulting in an underdetermined estimation 
    	problem.
In contrast, when $T=N=16$ and the switching-based training scheme described in 
Section~\ref{sec:LS} is employed, LS and LMMSE benefit from $N$ independent measurements, which 
substantially improves their accuracy, especially at moderate and high SNRs.}
	The learning-based approaches, 
	PAformer~\cite{PA_channel} and the proposed NN, maintain consistently superior NMSE performance across the 
	entire SNR range even without relying on switching (i.e., when only a single aggregated observation is available 
	per pilot slot). By exploiting the nonlinear mapping between the low-dimensional received pilot sequence and the 
	high-dimensional channel, the learning-based approaches are able to effectively infer the underlying spatial 
	structure of $\mathbf{h}_{u,k}$ from the input features and remain robust across the entire SNR range.  {However, the plain LSTM baseline exhibits significantly worse performance, indicating that a generic recurrent model is insufficient to fully capture the PA spatial structure without the proposed positional embedding and spatial modeling components.
	{The proposed LSTM-based architecture achieves slightly lower NMSE than PAformer and substantially outperforms the plain LSTM baseline across all SNR values, showing powerful channel estimation performance despite using simpler recurrent and dense operations rather than multi-head self-attention.}
		This efficiency stems from the combination of per-antenna temporal processing, global pilot-context extraction, position-aware Fourier features, and spatial BiLSTM modeling. Moreover, the architecture achieves this performance with significantly lower computational complexity and without the overhead of switching-based LS estimation.}

    {Fig.~\ref{Fig:Vary_SNR_t} shows the NMSE versus SNR performance for estimating the LoS component of the sensing channel (i.e., $\boldsymbol{\Theta}_{\mathrm{LoS}}$). {As expected, both LS and LMMSE exhibit limited estimation capability when $T=8$, while their performance improves when $T=16$ due to the availability of one measurement per activated PA.
    	{At very low SNR, however, the estimation error is dominated by noise, which can make the short-pilot and full-pilot classical baselines exhibit similar behavior.}
	It can be observed that the proposed LSTM-based estimator achieves substantially lower NMSE than LS, LMMSE, and the plain LSTM baselines, especially at moderate and high SNRs.
	The plain LSTM baseline achieves similar performance to the LMMSE, indicating that generic recurrent processing alone is insufficient for recovering the LoS sensing channel from clutter-contaminated scalar observations.}
    	In contrast, the ability to jointly leverage per-antenna temporal features and position-aware structure 
	allows the  \textcolor{black}{proposed LSTM-based estimator} to effectively learn the underlying LoS sensing channel even when $T<N$. The resulting performance gain is 
	especially pronounced at moderate and high SNRs, demonstrating the effectiveness of the proposed learning  framework under the observation constraints imposed by PA-enabled sensing.}

	\subsection{Impact of Number of Pilot Slots, $T$}
	
		\begin{figure}[!t]
		\centering
		\subfloat[]{
			\includegraphics[width=0.49\textwidth]{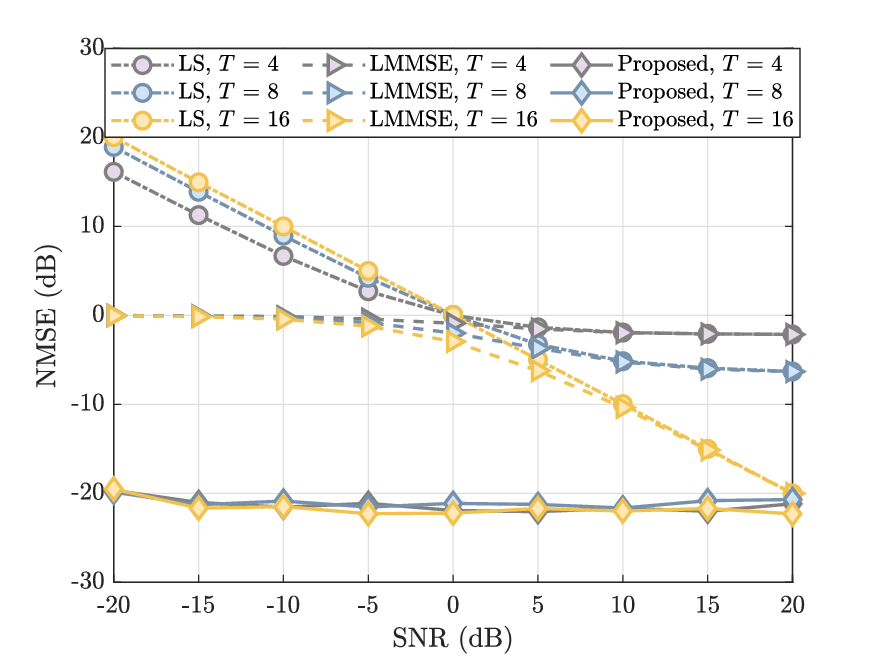}
			\label{Fig:Vary_T_u}
		}
		\subfloat[]{
			\includegraphics[width=0.49\textwidth]{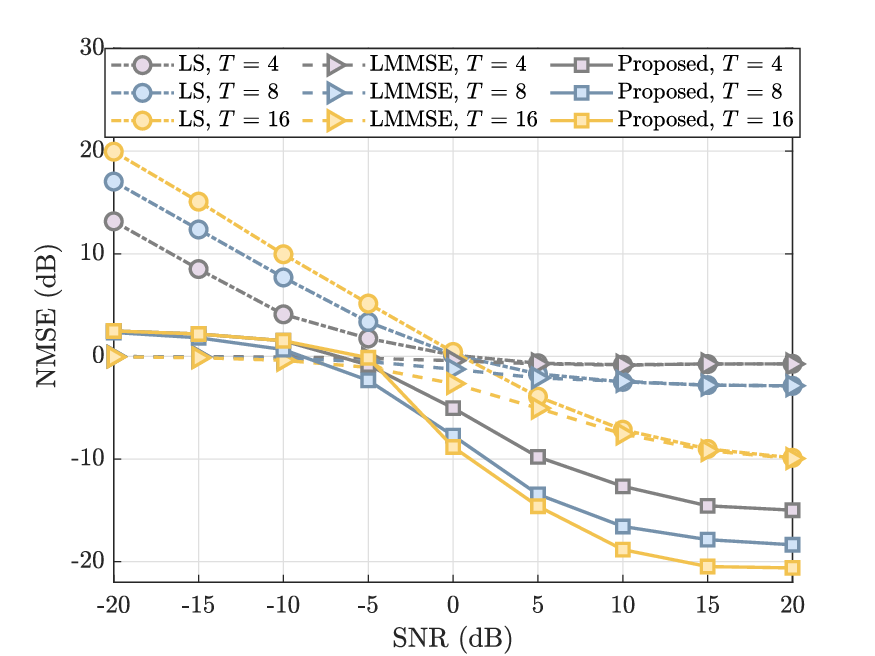}
			\label{Fig:Vary_T_t}
		}
		\caption{{NMSE performance versus SNR at $N = 16$. (a) Communication, (b) Sensing.} }
		\label{Fig:VaryT}
	\end{figure}
	
	{Fig.~\ref{Fig:VaryT} investigates the impact of the number of pilot slots on the channel estimation performance for $T\in\{4,8,16\}$ at $N=16$. To provide a practical interpretation of pilot overhead, the 5G NR frame structure can be used as a reference. According to~\cite{3gpp2016eutra}, one radio frame contains $80$ slots. Hence, the pilot overhead ratio is $\eta_{\rm p}=(T/80)\times100\%$, and the considered pilot lengths correspond to $5\%$, $10\%$, and $20\%$ overhead, respectively. As shown in Fig.~\ref{Fig:Vary_T_u}, the proposed estimator achieves stable communication channel reconstruction for all pilot lengths. Notably, even the short-pilot setting with $T=4$ attains performance comparable to the longer-pilot cases, demonstrating that the proposed model can effectively exploit the PA-induced spatial structure and pilot-domain dependencies from limited observations. In contrast, LS and LMMSE are more sensitive to the pilot length, especially when $T<N$, where the recovery problem becomes underdetermined. Fig.~\ref{Fig:Vary_T_t} shows the sensing-channel estimation performance. The sensing task is more challenging due to the bistatic target response and NLoS clutter, therefore, increasing $T$ improves the proposed estimator more noticeably than in the communication case. Nevertheless, the proposed method still substantially outperforms both LS and LMMSE under all pilot lengths. In particular, $T=4$ provides a very low-overhead operating point with $5\%$ pilot overhead, while $T=8$ offers improved reconstruction fidelity with only $10\%$ overhead. 
	Overall, this indicates that the proposed estimator can reduce the required pilot overhead by up to $75\%$ compared with LS and LMMSE, while maintaining reliable SAC channel reconstruction.}

	\subsection{\textcolor{black}{Impact of Varying the Number of PAs, $N$}}
	
			\begin{figure}[!t]
				\centering
				\subfloat[]{
						\includegraphics[width=0.49\textwidth]{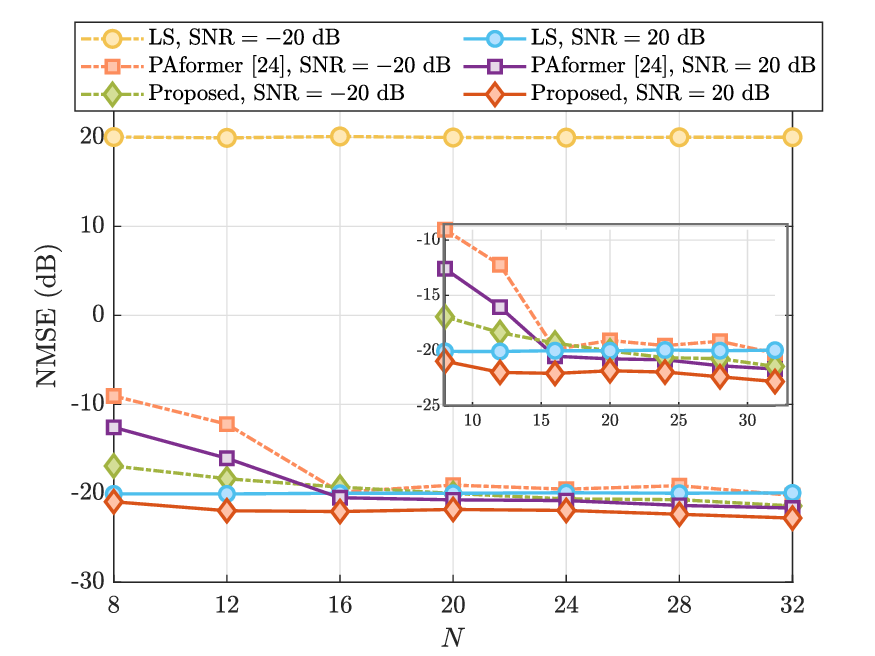}
						\label{Fig:Vary_N_u}
					}
				\subfloat[]{
						\includegraphics[width=0.49\textwidth]{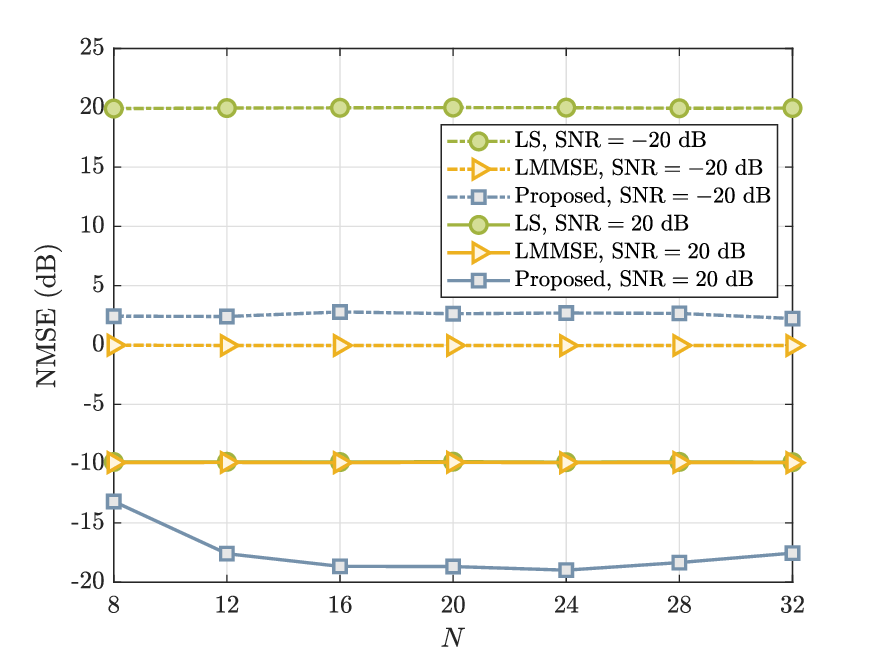}
						\label{Fig:Vary_N_t}
					}
				\caption{{NMSE performance versus $N$. (a) Communication, (b) Sensing.} }
				\label{Fig:VaryN}
			\end{figure}
			
			Fig.~\ref{Fig:VaryN} illustrates the impact of varying the number of activated PAs, $N$, on the NMSE 
			performance for both SAC links. Two representative SNR levels, $-20$~dB and $20$~dB, 
			are considered to highlight the behavior in noise-limited and {low-noise regimes, respectively.} The LS estimator is evaluated with $T=N$ pilot slots, whereas the learning-based methods are {trained and tested using 
			only $T=8$ pilots.} 
            
            Fig.~\ref{Fig:Vary_N_u} presents the {NMSE} results for the communication link. At low SNR (i.e., $-20$~dB), the LS estimator achieves higher NMSE than the learning-based approaches, consistent with the results from Fig~\ref{Fig:VarySNR}.  It is worth noting that LS remains  insensitive to the variation in $N$, since the switching scheme provides one independent measurement per PA. For the PAformer-based method, a mild degradation is observed for small $N$,  which can be attributed to the fact that the model was trained at $N=16$, and therefore encounters a distribution shift when evaluated at different antenna counts. In contrast, the proposed LSTM-based estimator maintains robust performance across all tested values of $N$ for 
			both SNR regimes. This behavior highlights the model's strong generalization capability: although trained only 
			at $N=16$, the architecture is able to process different antenna configurations without the need for retraining. This 
			property is particularly attractive for PA-assisted systems, where the number and placement of activated PAs 
			may vary dynamically depending on user locations or service requirements. The consistent NMSE trends across 
			$N$ confirm that the proposed model effectively adapts to such variations. 
			{Fig.~\ref{Fig:Vary_N_t} shows the corresponding NMSE performance for the sensing link. Similar to the communication case, the LS and LMMSE estimators remain almost invariant with respect to $N$ However, the LS  NMSE performance is substantially higher at low SNR. In contrast, the proposed LSTM-based estimator maintains stable performance across all values of $N$ and consistently outperforms both LS and LMMSE at high SNR by a considerable margin. The robustness to variations in the number of activated PAs further confirms the strong generalization capability of the proposed architecture, which is particularly beneficial in PA-assisted sensing scenarios where $N$ may vary depending on operational requirements.}
			
	
	\subsection{Impact of Number of Scatterers, $S$}
	     \begin{figure}
		\centering
		\subfloat[]{
			\includegraphics[width=0.49\textwidth]{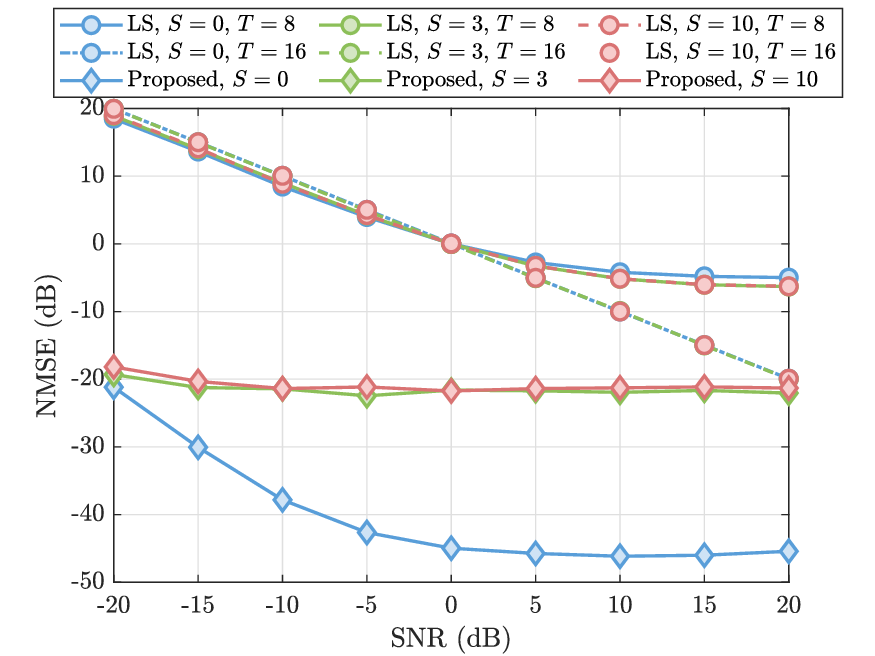}
			\label{Fig:SUinv}} 
		\subfloat[]{
			\includegraphics[width=0.49\textwidth]{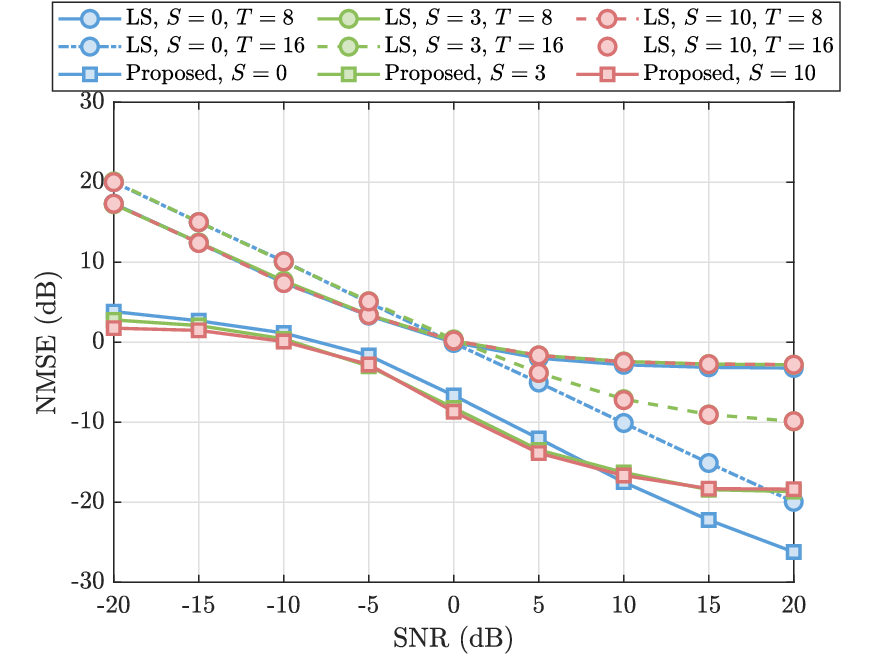}
			\label{Fig:STinv}}
		\caption{{Impact of the number of scatterers, $S$, on NMSE performance. (a) Communication, (b) Sensing.}}
		\label{Fig:S}
	\end{figure}
		
		{Fig.~\ref{Fig:S} evaluates the impact of the number of environmental scatterers on the channel estimation performance by varying $S\in\{0,3,10\}$. For the LS estimator, both $T=8$ and $T=16$ are considered to distinguish the effect of limited pilot observations from the effect of scattering. For the proposed estimator, $T=8$ is considered to highlight its performance under compressed pilot observations. As shown in Fig.~\ref{Fig:SUinv}, the communication-channel estimation performance of LS is mainly affected by the pilot length rather than the number of scatterers. This is because the desired communication channel includes both LoS and NLoS components; hence, the scatterers are part of the effective channel to be estimated rather than interference. The proposed estimator achieves robust performance for both $S=3$ and $S=10$, while the case $S=0$ exhibits strong reconstruction accuracy by eliminating random scattered components and enabling a more direct mapping between the PA-dependent observations and the effective channel. Fig.~\ref{Fig:STinv} shows the sensing-channel estimation performance. Unlike the communication case, the NLoS components act as clutter. Therefore, increasing $S$ makes the sensing estimation task more challenging. The LS estimator suffers from noticeable degradation in cluttered scenarios, especially at high SNR where the clutter-induced mismatch remains. In contrast, the proposed estimator consistently outperforms LS for all considered values of $S$. These results show that the proposed model can reliably recover the effective communication channel under different scattering conditions.}
		

	\subsection{Robustness to PA Placement Errors}
	  \begin{figure}[!t]
		\centering
		\subfloat[]{
			\includegraphics[width=0.49\textwidth]{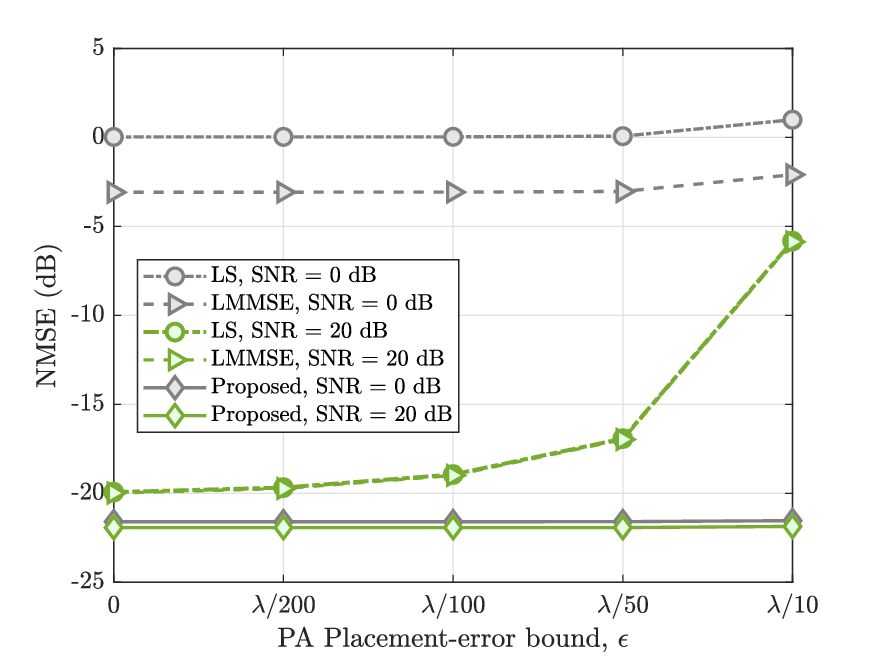}
			\label{Fig:userPAerror}}
		\subfloat[]{
			\includegraphics[width=0.49\textwidth]{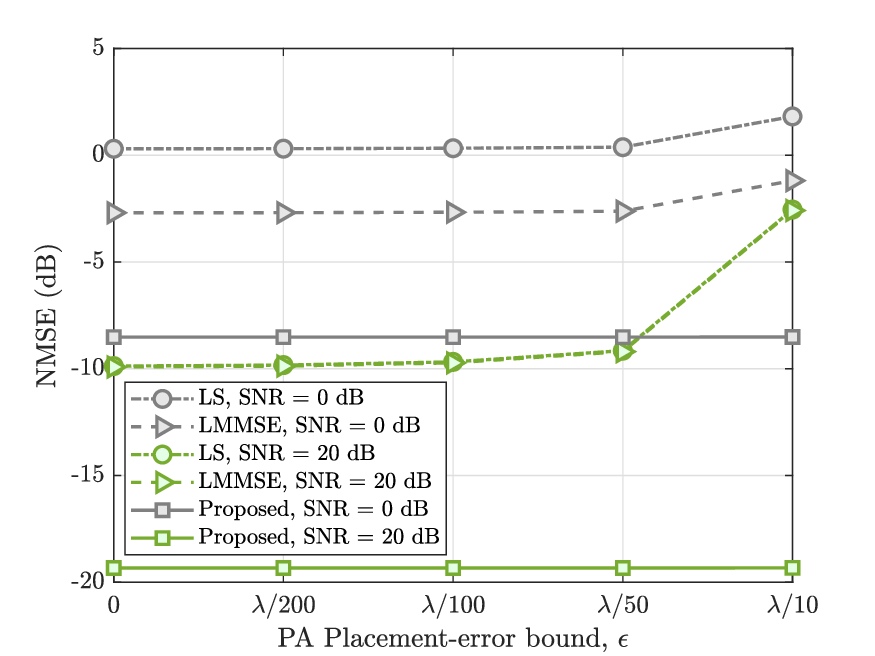}
			\label{Fig:targetPAerror}}
		\caption{{NMSE performance versus PA placement-error bound, $\epsilon$. (a) Communication, (b) Sensing. }}
		\label{Fig:PAError}
	\end{figure}
	
	
	{To evaluate the sensitivity of the positional embeddings to practical PA placement mismatch, a bounded perturbation is introduced to the PA position features during testing. The training stage is performed using the true PA positions, while the input to the Fourier positional embedding  is perturbed according to $
	\hat{x}_n = x_n + \Delta x_n, 
	\Delta x_n \sim \mathcal{U}[-\epsilon,\epsilon],$ where $\epsilon$ denotes the PA placement-error bound. The considered values are $\epsilon \in \{0,\lambda/200,\lambda/100,\lambda/50,\lambda/10\}$. Fig.~\ref{Fig:PAError} shows the resulting NMSE performance for both communication and sensing links. As observed in Fig.~\ref{Fig:userPAerror}, the proposed  estimator remains nearly unchanged across all considered placement-error bounds, indicating strong robustness to coordinate mismatch. In contrast, the LS and LMMSE estimators degrade noticeably as $\epsilon$ increases, particularly at high SNR where the mismatch error becomes dominant. A similar behavior is observed for the sensing link in Fig.~\ref{Fig:targetPAerror}, where the proposed estimator maintains stable NMSE performance, while the classical estimators exhibit clear sensitivity to PA placement errors. These results demonstrate that the Fourier-based positional embeddings preserve reliable performance under small practical PA placement mismatches.}
	
	\subsection{MUSIC Estimation Performance}

             {Fig.~\ref{Fig:MUSIC} illustrates the target parameter estimation accuracy obtained by the 
			2D near-field MUSIC algorithm described in Section~\ref{Sec:Music} based on the estimated LoS sensing channel, $\hat{\boldsymbol{\theta}}_{\mathrm{LoS}}$.
		{Specifically, Fig.~\ref{Fig:MUSIC_aoa} compares the AoA RMSE versus SNR of the proposed LSTM-based estimator with those of LS and LMMSE.}
			The proposed estimator achieves remarkably accurate angle recovery, reaching an AoA RMSE on the order of $10^{-2}$ degrees  at $\mathrm{SNR}=-15$~dB, despite relying on only $T=8$ pilot symbols.
		{In contrast, the LS and LMMSE estimators with $T=8$ yield severely degraded results across all SNR values due to their poor channel estimates, causing MUSIC to operate with noise-dominated spatial signatures.
			With $T=N=16$, both LS and LMMSE attain reasonable accuracy at higher SNRs, yet their RMSEs remain substantially larger in the low-SNR regime.}
			These observations confirm that the channel estimates produced by the proposed LSTM model preserve the  geometric structure  for subspace methods and enable highly accurate angle estimation under  various pilot and noise conditions.
			
			Fig.~\ref{Fig:MUSIC_r} presents the RMSE {versus SNR performance corresponding to target range.} A similar trend is observed {as in Fig.~\ref{Fig:MUSIC_aoa}}. The 
			proposed \textcolor{black}{LSTM-based} estimator provides precise range predictions even at low SNR, closely tracking the true range for SNRs above $-15$~dB. 
			{In contrast, LS and LMMSE with $T=8$ are unable to recover the range, where their severely biased channel estimates lead to pronounced error floors. However, with $T=16$, both baselines achieve acceptable accuracy only at moderate and high SNR levels.
			 These results highlight that the proposed estimator better preserves the spherical-wave propagation characteristics necessary for reliable 2D MUSIC processing.}
			 

			\begin{figure}[!t]
				\centering
				\subfloat[]{
					\includegraphics[width=0.49\textwidth]{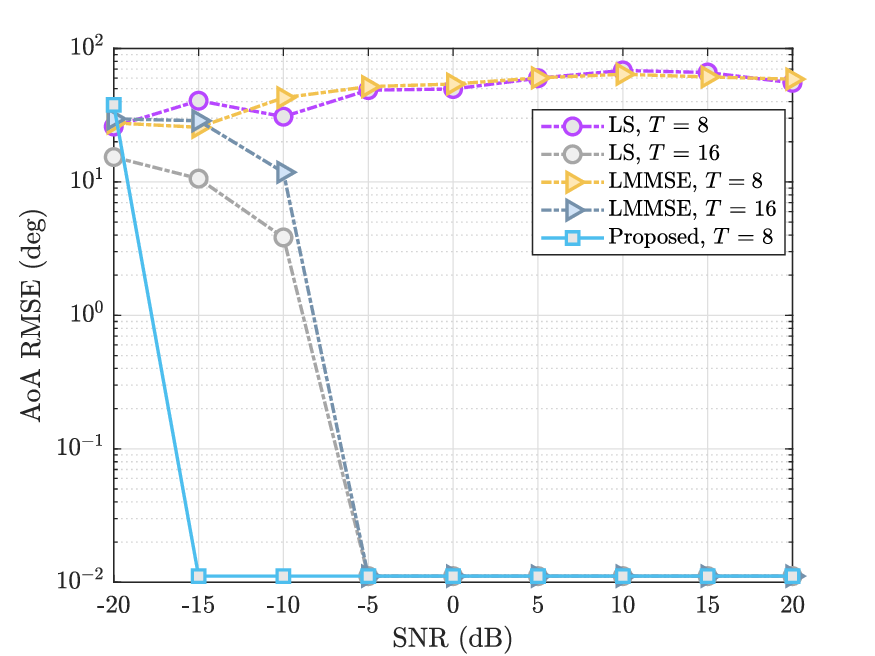}
					\label{Fig:MUSIC_aoa}
				}
				\subfloat[]{
					\includegraphics[width=0.49\textwidth]{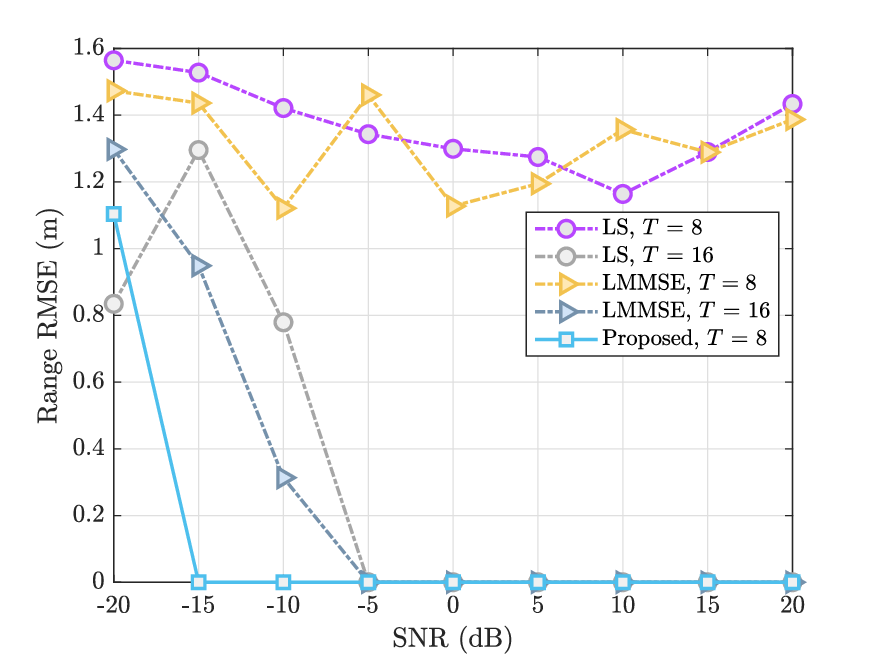}
					\label{Fig:MUSIC_r}
				}
				\caption{{Target parameter estimation performance versus SNR. (a)  AoA, (b) Range. }}
				\label{Fig:MUSIC}
			\end{figure}

				Furthermore, to illustrate the behavior of the MUSIC estimator based on the proposed LSTM LoS channel estimates, Fig.~\ref{Fig:MUSIC_S} depicts the MUSIC angle pseudospectrum $P(r\,|\,\varphi)$ at several representative SNR levels {$\in\{-20,-15,15,20\}$} dB. Each {subfigure} corresponds to the angle slice of the {2D} near-field MUSIC spectrum evaluated at the estimated range $\hat r$. At very low SNR (e.g., $\text{SNR}=-20$~dB), the inaccuracy of the channel estimate  leads to a false range estimate $\hat r$, which misaligns the steering vector with the true bistatic geometry. As a result, the  dominant peak does not appear at the true angle. As the SNR increases to $-15$~dB and above, the LSTM-based estimator provides sufficiently accurate LoS structure, enabling MUSIC to correctly localize the target. In these cases, a sharp and isolated peak emerges at the true direction $\varphi \approx 20^\circ$, while sidelobes remain well suppressed. At high SNR values ($15$--$20$~dB), the peak sharpens further and the spectrum stabilizes, confirming that the estimated channel preserves the spatial phase profile required for reliable  parameter extraction.
				
					\begin{figure}[!t]
			\centering
			\includegraphics[width=0.65\textwidth]{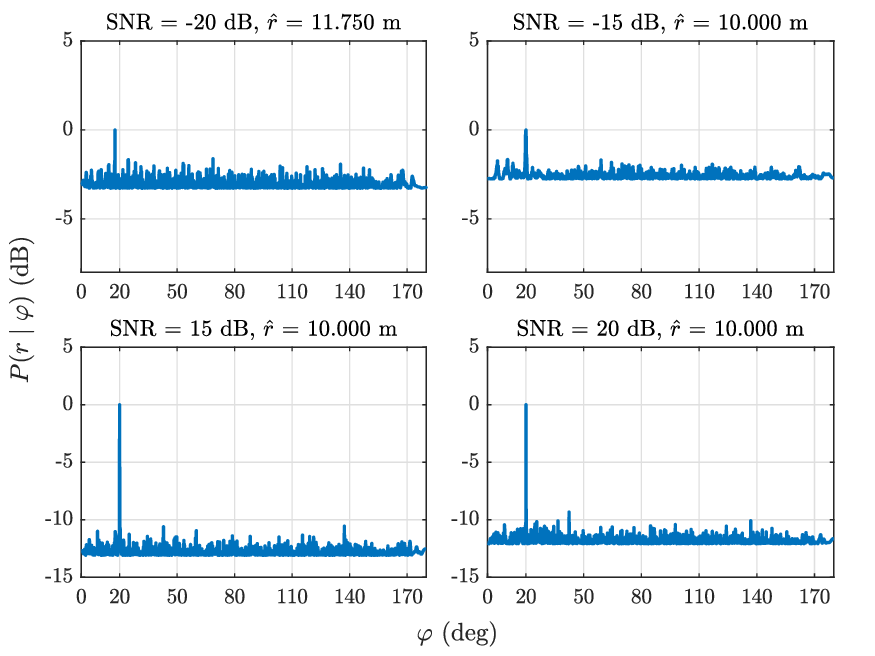}
			\caption{MUSIC angle pseudospectrum at different SNR levels using LSTM-based channel estimates. \vspace{-7mm}}
			\label{Fig:MUSIC_S}
		\end{figure}
		

	\subsection{Complexity Evaluation}

	Fig.~\ref{Fig:Comp} presents the computational complexity of the proposed LSTM-based 
	estimator for the communication channel $\mathbf{h}_{u,k}$, expressed in terms of the number of real 
	multiplications and additions derived in Section~\ref{Sec:Complexity_Analysis}. The complexity is compared 
	against the PAformer architecture in~\cite{PA_channel} for a wide range of activated PAs $N$. 
	It can be observed that the proposed \textcolor{black}{LSTM-based estimator} achieves a substantial reduction in computational cost, 
	with the overall complexity decreasing by approximately $56\%$--$62\%$ relative to PAformer as $N$ 
	increases. This gain arises from the lightweight sequential structure of the LSTM, which avoids the 
	quadratic attention operations inherent to transformer-based designs. Remarkably, these savings are 
	achieved while delivering slightly better channel-estimation accuracy, as demonstrated earlier in 
	Figs.~\ref{Fig:VarySNR} and \ref{Fig:VaryN}. Therefore, the proposed solution offers both improved estimation fidelity and significantly enhanced computational efficiency, making it highly attractive for practical PA-assisted ISAC systems.

	\begin{figure}[!t]
		\centering
		\includegraphics[width=0.6\textwidth]{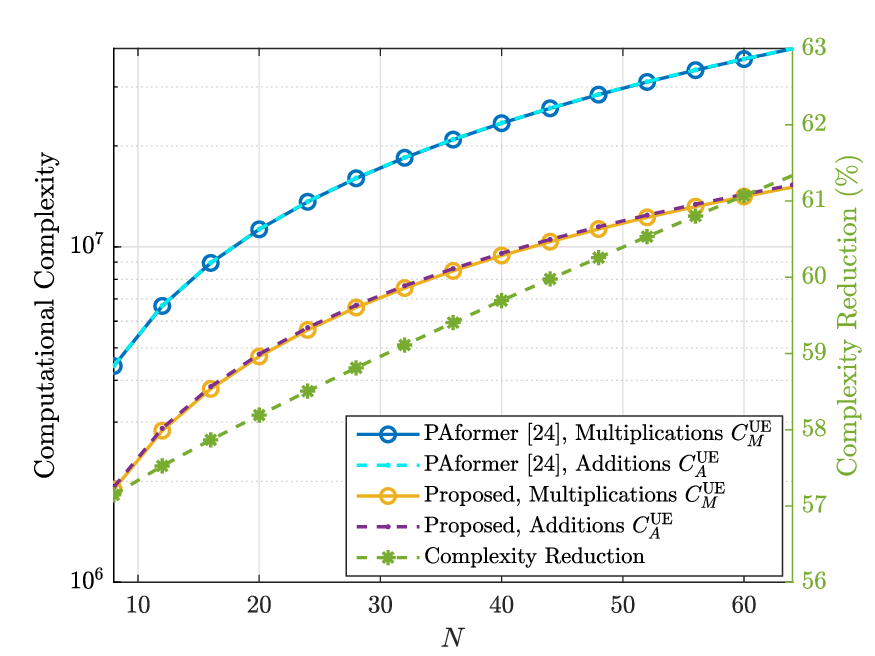}  
		\caption{Computational complexity evaluation for the communication link.}
		\label{Fig:Comp}
	\end{figure}

\section{Conclusion} ~\label{Sec:Conclusion}

{This paper investigated the channel estimation for a PA-enabled ISAC system and proposed a novel LSTM-based learning framework to recover SAC channels from highly compressed pilot measurements. By exploiting the sequential structure of received signals and tailored feature representations, the proposed approach effectively addressed the low-rank characteristics of PA-based architectures and enabled reliable estimation in short-pilot regimes where classical methods struggle. Additionally, to extract sensing information, a near-field 2D MUSIC algorithm is integrated to estimate target angle and range using the LSTM-based LoS sensing channel. Extensive simulations across various SNRs and array configurations demonstrate the robustness and generalization capability of the proposed framework. Complexity analysis further showed that the LSTM model reduces computational cost by approximately $56$--$62\%$ compared to transformer-based PAformer architectures while achieving slightly better estimation accuracy, making it a practical and efficient solution for PA-enabled ISAC systems. {Future works can consider time-varying PA-enabled ISAC channels by incorporating Doppler-dependent temporal phase evolution across coherent sensing snapshots. In this direction, the proposed LSTM-based framework can be further extended to support joint channel, range, angle, and velocity estimation in high-mobility scenarios.}}

\bibliographystyle{IEEEtran}

{
	\bibliography{IEEEabrv,References}
}
\end{document}